\documentclass[12pt]{iopart}
\usepackage{iopams}
\usepackage{graphicx}

\def\tr{\mathop{\rm Tr}\nolimits}

\def\1{{\bf 1}}

\def\ket #1{\vert #1\rangle}
\def\braket #1#2{\langle #1 \vert #2\rangle}
\def\ketbra #1#2{\vert #1\rangle \langle #2\vert}

\begin{document}

\review[Quantum channels]{Quantum channels and their entropic characteristics}

\author{A S Holevo$^1$ and  V Giovannetti$^2$}

\address{$^1$Steklov Mathematical Institute, Gubkina 8, 119991 Moscow, Russia}
\ead{holevo@mi.ras.ru}
\address{$^2$NEST, Scuola Normale Superiore and Istituto Nanoscienze-CNR, piazza dei Cavalieri 7, I-56126 Pisa, Italy}
\ead{v.giovannetti@sns.it}

\begin{abstract}
One of the major achievements of
the recently emerged quantum information theory is
the introduction   and thorough investigation of the
notion of quantum channel which
is a basic building block of any data-transmitting or data-processing system.
This development resulted in an elaborated structural theory and was accompanied by the discovery of
a whole spectrum of entropic quantities, notably the channel capacities, characterizing
information-processing performance of the channels.

This paper gives a survey of the main
properties of quantum channels and of their entropic characterization,
with a variety of examples for finite dimensional  quantum systems.
We also touch upon the ``continuous-variables''
case, which provides an arena for quantum Gaussian systems. Most of
the practical realizations of quantum information processing were implemented
in such systems, in particular based on principles of
quantum optics.
Several important entropic quantities are introduced and
used to describe the basic channel capacity formulas.
The remarkable role of the specific quantum correlations --
entanglement -- as a novel communication resource, is stressed.
\end{abstract}

\pacs{03.67.-a, 05.30.-d}
\submitto{\RPP}

\maketitle

\tableofcontents

\newpage

\section{Introduction}

The concept of communication channel and its capacity is central in
information theory~\cite{cover} where probabilistic models of noise were used
with great success. The importance of considering quantum channels
is due to the fact that any physical communication line is after all a quantum
one, and it can be treated classically only in so far as the quantum
noise associated with  the fundamental quantum-mechanical uncertainty~\cite{dir58} is
negligible as compared to the classical fluctuations. This is not
the case in many modern applications such as optical communication~\cite{cd}
or quantum system engineering~\cite{nc}, calling for a genuinely quantum
approach.

Mathematically, the notion of quantum channel is related to that of
{\it dynamical map}, which emerged in 1960-70s  and goes back to
the ``operations'' of Haag-Kastler \cite{haa64} and Ludwig \cite{lud83}.
The essential property of dynamical maps is {\it positivity}
ensuring that states are transformed into states in the
Schr\"odinger picture. These transformations   are a quantum analog
of the Markov maps (stochastic matrices) in probability theory and a
natural further development was to consider semigroups of such processes
as a model of quantum Markovian dynamics (see Kossakowski \cite{kos72}, Davies \cite{d2}).

However, an approach based on positivity remained not extremely
productive until it was observed independently by several
researchers (e. g. Kraus \cite{kr}, Lindblad \cite{lin75}, Gorini et
al. \cite{gor78}, Evans and Lewis \cite{eva77} in statistical
mechanics, and by Holevo \cite{hol72} in the context of quantum
communication theory) that the reduced dynamics of open quantum systems
has a stronger property of  {\it complete positivity}, introduced
earlier in the purely mathematical context by Stinespring
\cite{stine} and studied in detail in the finite dimensional  case by Choi~\cite{choi}.
Moreover when interpreted in the physical terms, the basic Stinespring's dilation theorem
implies that complete positivity is not only necessary, but also sufficient for a dynamical map to be
extendable to the unitary dynamics of an open quantum system
interacting with an environment. In this way the notion of dynamical
map in statistical mechanics (and that of quantum channel in information
theory) was finally identified with that of completely positive, properly normalized
map acting on the relevant operator space (generated either by
states or by observables) of the underlying quantum system. A comprehensive
study of the dynamical aspects of open quantum systems was presented
by Spohn \cite{spo80}, and a number of important entropic quantities
introduced already at this early stage were surveyed by Wehrl
\cite{wehrl}.

Quantum information theory brought a new turn to the whole subject
by giving a deeper operational insight to the notion of channels and to their
 entropic characteristics. Fundamental  results of the classical
information theory are the {\it coding theorems} which establish the
possibility of  transmitting  and processing data reliably (i.e.  asymptotically error-free)
at rates not exceeding  certain threshold  values ({\it capacities})
that characterize the system under consideration. Coding
theorems provide explicit formulas for such thresholds in terms of
entropic functionals of the channel. From a different
perspective, quantum information theory gives a peculiar view onto
irreversible evolutions of quantum systems, providing the
quantitative answer to the question -- to what extent effects of
irreversibility and noise in the channel can be reversed by using
intelligent pre-processing of the incoming and post-processing of
the outgoing states. The issue of the information capacity of
quantum communication channels arose soon after publication of the
pioneering Shannon's paper \cite{shannon} and goes back to works of
Gabor \cite{gabor} and Gordon \cite{gor}, asking for fundamental
physical limits on the rate and on the quality of the information transmission.
This laid a physical foundation and raised the question of
consistent quantum-theoretical treatment of the problem.

Important steps in this direction were made in the 1970-s when
quantum detection and estimation theory was developed, making a
quantum probabilistic frame for these problems, see
the books of Helstrom \cite{he} and Holevo \cite{ho1}. At that time the concepts of
quantum communication channel and of its capacity for transmitting
classical information were established, along with a fundamental
upper bound for that quantity. From this point the subject of the
present survey begins. In 1990-s a new interest on
noisy quantum channels arose in connection with the emerging quantum
information science, with a more detailed and deeper insight -- e.g. see
the books~\cite{nc,hayashi} or the review papers~\cite{bs,wer,s,KEYL} and the references therein
(further references to original works will be also given in the present reivew).
The more recent developments are characterized by an emphasis on the new possibilities (rather than
on the restrictions) implied by specifically quantum features of the
information processing agent, notably -- the {\it entanglement} as a novel communication resource.
Along these lines the achievability of the information bound and a
number of further quantum coding theorems were discovered making
what can be described as {\it quantum Shannon theory}. It was
realized that a quantum communication channel is characterized by a
whole spectrum of capacities depending on the nature of the
information resources and specific protocols used for the
transmission. On the other hand, the question of information
capacity turned out important for the theory of quantum computing,
particularly in connection with quantum error-correcting codes,
communication and algorithmic complexities, and quantum cryptography
where the channel environment not only introduces the noise but also
models an eavesdropper interfering with private communication.

 The aim of the present paper  is to provide a survey of the principal features of quantum communication channels.
We will discuss mathematical representations for quantum channels and focus on the question -- how their efficiency in
transferring signals can be characterized in terms of various entropic quantities.
The material is organized as  follows.
Sec.~\ref{sec:ChannelsOpen} is devoted to the
structural theory of quantum channels. It also gives a survey of the
main particular classes of channels in the case of finite level quantum
system with a variety of examples. In Sec. \ref{gaussianchannels} we pass to the
``continuous-variables'' case, which provides an arena for quantum
Gaussian systems. Many experimental demonstrations of quantum
information processing were realized in such
systems, in particular basing on principles of quantum optics.
Several important entropic
quantities are introduced in Sec.~\ref{entropic} and used to describe the basic
capacity formulas. Most of content in this Section is devoted to the finite level systems and
can be addressed straight after reading Sec.~\ref{sec:ChannelsOpen}. In the course of our survey
several important open questions are formulated and discussed. Summary and outlook are given in Sec.~\ref{sec:conclusion}.

\section{Channels and open systems}\label{sec:ChannelsOpen}

\subsection{Classical and quantum information carriers}\label{sec:ClassicalQuantum}

 In any information-processing scheme,
 data are encoded into the
states of some physical systems (e.g. the etched surface  of a DVD or the EM microwaves emitted by a cellphone)
which play the role of {\it information carriers} or fundamental data-processing elements.
In its simplest version, the classical theory of information
assigns to each of them  a finite phase space $\mathcal{X}$ (called also {\it alphabet}) whose elements  represent the possible configurations  the carrier can assume  (e.g. on/off, dot/dash/space in Morse code, etc.).
More precisely  the completely determined states  ({\it pure states}) of a carrier
are identified with  the points $x \in \mathcal{X}$, while
the  {\it mixed  states} which represent statistical ensemble of
pure states,  are described by probability distributions
$\{p_x\}$  on~$\mathcal{X}$.
The Shannon entropy of such distribution~\cite{cover}
\begin{eqnarray} \label{defshannon}
H(X)=H(\{ p_x\})  \equiv -\sum_{x}p_{x}\log _{2}p_{x} \;,
\end{eqnarray}
provides a measure of the ``uncertainty" or of the  ``lack of information" in the
corresponding ensemble (also, as it will be clarified in Sec.~\ref{transclas}, $H(X)$
describes the potential information
content of a {\it random source} producing symbols $x$ with
probabilities~$p_x$).
Choosing the binary logarithm in Eq.~(\ref{defshannon}) means that we are measuring  information in
binary digits -- {\it bits} -- which is convenient because of the basic role
of two-state  processing systems.
 The minimal value of the entropy, equal to~$0$, is attained on pure states  (here and in the remaining of
the paper we adopt the usual convention $0\log_20=0$), while the maximal,
equal to~$\log_2 |\mathcal{X}|$, is obtained on the uniform distribution
$p_{x}\equiv |\mathcal{X} |^{-1}$, where $|\mathcal{X}|$ denotes the
size of the alphabet $\mathcal{X}$.

Such a simple   phase space description does not hold when
the information carrier used to encode the data is  a quantum system
(e.g.  a two-level ion  confined in space by  using strong electromagnetic
fields~\cite{MONROE} or a  single polarized photon propagating along an optical fiber~\cite{GISIN}). Instead the latter  can be represented in terms
of a Hilbert space $\mathcal{H}$~\cite{dir58,neu32}, which again for simplicity we take
finite dimensional. In this context, the  {\it pure quantum states}  are described by
projections $|\psi \rangle \langle \psi |$ onto unit vectors $|\psi
\rangle $ of $\mathcal{H}$ (as in the classical case they define the completely determined configurations of the carrier, which in the quantum case correspond
to specific state preparation procedures).
 {\it Mixed states} are
represented by  statistical ensembles of pure states
$|\psi^{\alpha}\rangle \langle \psi^{\alpha}|$ with probabilities $p_\alpha$. Each mixed state is
formally described by the corresponding {\it density operator} $\rho \equiv \sum_{\alpha}p_{\alpha}|\psi^{\alpha}\rangle \langle \psi^{\alpha}|$ which
contains,  in a highly condensed form, the information about the procedure for producing such state
(specifically  it describes a preparation procedure of the carrier characterized by a stochastically fluctuating  parameter~$\alpha$),  and which has the
following properties~\cite{neu32}:
\begin{itemize}
 \item[{\it i)}] $\rho$ is a linear Hermitian
positive operator in ${\cal H}$;
\item[{\it ii)}]  $\rho$ has unit trace, i.e. $\mbox{Tr} \rho =1$.
\end{itemize}
Analogously to their classical counterparts, quantum mixed states form a convex set~$\mathfrak{S} ({\cal H})$ whose extremal
points are represented by pure states.
Unlike classical mixed states however,  they  can be written as
mixture of pure states in many different ways, i.e. one and the same density matrix $\rho$
represents different stochastic preparation procedures of the carrier.  A
distinguished ensemble is given by the spectral decomposition
$\rho =\sum_{j}\lambda_{j}\;|e^{j}\rangle \langle e^{j}|$,
where $\lambda_{j}$ are the eigenvalues and $|e^{j}\rangle $ --
orthonormal eigenvectors of the operator $\rho .$ The
eigenvalues $\lambda_{j}$ of a density operator form a probability
distribution and the Shannon entropy of this distribution is
equal to the von Neumann entropy of the density operator~\cite{neu32},
\begin{eqnarray} \label{NEUMANN}
S(\rho ) \equiv -\mbox{Tr} \rho \log _{2}\rho =-\sum_{j}\lambda _{j}\log
_{2}\lambda _{j}= H(\{\lambda_j\} )\;,
\end{eqnarray}
which provides a measure of uncertainty and, as it is explained
later, of the information content of the quantum state $\rho$. Again, the
minimal value of the entropy, equal to $0$, is attained on pure states
while the maximal, equal to $ \log_2 d$, -- on the
{\it chaotic state} $\rho = I/d$ (with
$I$ denoting the unit operator in $\mathcal{H}$ and $d=\dim \mathcal{H}$ being the dimensionality of the space).

There is a way to formally embed a finite classical system associated
with a classical information carrier into a quantum system by
introducing the Hilbert space with the orthonormal basis $\left\{
|x\rangle \right\} $ indexed by phase space points $x\in
\mathcal{X}.$ Then to the classical states $\{p_x\}$ of the
classical carrier correspond the diagonal density operators
$\rho =\sum_{x}p_{x}|x\rangle \langle x|$
which commute with each other.  Truly quantum systems, however,
admit also configurations which
are described by density operators that are {\it not} simultaneously
diagonalizable (i.e. noncommuting).
Indeed the full information content of a quantum state cannot be reduced to a
classical message and therefore deserves special name
{\it quantum information}. This is related to the fact that
a quantum state contains implicitly the statistics of all possible
quantum measurements, including mutually exclusive (complementary)
ones: a distinctive feature which ultimately leads to  the
{\it impossibility of cloning } of quantum information (see Sec.~\ref{irreversibility} for detail).
\newline

{\bf Example: }  Take a binary alphabet ${\cal X}=\{ 0,1\}$ with associated
quantum orthogonal basis $\{|0\rangle, |1\rangle\}$. The  density operator  $|+\rangle\langle +|$ corresponding to the vector  $|+\rangle=(|0\rangle + |1\rangle)/\sqrt{2}$,
is a proper quantum state of the system  which  in general does not commute with the density matrices $p_0 |0\rangle \langle 0 | + p_1 |1\rangle\langle 1|$
 used to  embed the statistical distributions $\{ p_0,p_1\}$ on ${\cal X}$.
\newline

In information theory, both classical and quantum, messages are typically
transmitted by using block coding strategies that exploit long sequences of carriers. Therefore one systematically has
to do with composite systems corresponding to repeated or parallel
uses of communication channels. {\it Entanglement} reflects
unusual properties of composite quantum systems which are described
by tensor rather than Cartesian (as in the classical case) product
of the component systems. According to  the superposition
principle~\cite{dir58}, the  Hilbert space ${\cal H}_{AB} \equiv {\cal H}_A \otimes {\cal H}_B$
of the composite system $AB$,
along with product vectors
$|\psi_A\otimes \psi_B\rangle \equiv |\psi_{A}\rangle \otimes |\psi _{B}\rangle$ with
 $|\psi_A\rangle\in{\cal H}_A$ and $|\psi_B\rangle\in{\cal H}_B$,
 contains all possible
linear combinations $\sum_{\alpha}|\psi _{A}^{\alpha} \otimes \psi
_{B}^{\alpha}\rangle $. The pure states given by product vectors are
called {\it separable} while their superpositions are
{\it entangled}. A mixed state of $AB$ is called separable if it
can be expressed as a mixture of product states, and entangled if it
can not~\cite{ENTANGLEMENT,nc}.

Entanglement is an intrinsically  ``nonclassical''   sort of correlation between
subsystems which typically emerges due to quantum interactions.
If a classical composite system $AB$ is in a pure (i.e. completely
determined) state then apparently the subsystems $A,B$ are also in their
uniquely defined pure partial states. Strikingly, this is not so for the entangled
states of a quantum composite system $AB$. Consider a generic
pure state $|\psi _{AB}\rangle$
of such a  system. By {\it Schmidt decomposition} it can be expressed as
 \begin{eqnarray}
\ket{\psi_{AB}}=\sum_{j}\sqrt{\lambda _{j}}\; \ket{e^j_A \otimes e^j_B}\;,
\label{sch}
\end{eqnarray}
where  $\left\{ \lambda _{j}\right\} $  is a probability distribution which is uniquely determined by $|\psi _{AB}\rangle$,  while $\{|e_{A,B}^j\rangle\}$ are some orthonormal bases in $\mathcal{H}_{A,B}$~\cite{nc}.
Then the local  state of the subsystem $A$ is  given by the {\it reduced}  density operator
\begin{eqnarray}
\rho _{A}&=&\mbox{Tr}_B \ket{\psi_{AB}} \langle \psi_{AB}|= \sum_{j}\lambda _{j}\ketbra{e^j_A}{e^j_A}\;,
\label{part}
\end{eqnarray}
obtained  by taking the partial trace $\mbox{Tr}_B[\cdots]$ of the
original joint state~(\ref{sch}) over the degrees of freedom of $B$
(similarly for the partial state of $B$). Thus, unless the Schmidt  form~(\ref{sch}) factorizes (i.e. the numbers $\lambda_j$ are all but one equal to zero),
 the density matrices of $A$ and $B$ represent mixed states. In particular
they have the same nonzero eigenvalues $\lambda _{j}$
and hence  equal entropies,
\begin{eqnarray}
S(\rho _{A})=S(\rho _{B})=-\sum_{j}\lambda _{j}\log_2 \lambda _{j}\;,
\label{eent}
\end{eqnarray}
which are strictly positive if the state  $|\psi _{AB}\rangle $ is a
genuine superposition of product vectors,~i.e. entangled.
In fact, the entropy (\ref{eent}) measures how much entangled is the state~(\ref{sch});
it is zero if and only if $\ket{\psi_{AB}}$  is separable, and
takes its maximal value $\log_2 d$ for the {\it maximally
entangled} states
which have the Schmidt representation~(\ref{sch}) with the uniform coefficients
$\lambda_j  = 1/\sqrt{d}$ (measures of entanglement  can also be
defined for mixed states of $AB$ -- we refer the reader
to~\cite{HHHH,BRUS,GUNE} for detailed reviews on that subject).

This passage from the pure joint state of a composite system  to the  mixed partial state
of one of its components  can be inverted.
Indeed,  for any density matrix $\rho_A$ of  a system $A$
there is a pure state of a composite system $AB,$ where $B$ is large
enough to contain a copy $A,$ given by the vector (\ref{sch}) such
that $\rho _{A}$ is its partial state. This pure state is called
{\it purification} of $\rho _{A}.$ Nothing like this exists in
the classical system theory.

For the sake of completeness it is finally worth to mention that
entanglement is not the sole way in which  quantum mechanical states
exhibit nonclassical features. Another aspect which
 is currently attracting a growing interest
  is the possibility of exploiting the nonorthogonality of states,
  like $|0\rangle$ and $|+\rangle$ introduced before,
  to build correlated joint configurations which, even though being separable,
  cannot be accounted for  by  a purely classical theory.
A measure for such
nonclassical, yet unentangled, correlations  is provided
by the so called {\it quantum discord},~see \cite{OLLI,HENDE} for detail.
\newline

\subsection{Classical and quantum channels}\label{Sec22}

 A communication channel is
 any physical transmission medium (e.g. a wire) that allows two parties (say, the sender and the receiver) to
 exchange messages.  In classical information theory~\cite{cover},  a channel is abstractly modeled by
 specifying an input alphabet $\mathcal{X}$  and  an output alphabet $\mathcal{Y}$
(with $\mathcal{Y}$ not necessarily the same as $\mathcal{\ X}$).
 The elements $x$ of $\mathcal{X}$ represent the input signals (or {\it letters})
the sender wishes to transfer. Alternatively one can think of $x$'s
as the pure (classical)  states at the input of the information carrier that propagate through
 the physical medium that describes the communication line.
 Similarly the elements $y$ of $\mathcal{Y}$ represent the
output counterparts of the input signals which arrive to the
receiver after the propagation through the medium. The physical
properties of the communication process, including the noise that
may affect it,  are then summarized by a conditional
probability $p(y|x)$ of receiving an output letter $y\in
\mathcal{Y}$ when an input signal $x\in \mathcal{X}$ is sent. Hence,
if an input probability distribution $P=\{p_{x}\}$ is given,
reflecting the frequencies of different input signals, then the
input and the output become random variables $X,Y$ with the joint
probability distribution $p_{x,y}=p(y|x)p_{x}.$ Accordingly, the
input probability distribution $P=\{p_{x}\}$ is transformed by the
channel to the output probability distribution $P^{\prime}=\{p_{y}^{\prime }\}$,
 where $p_{y}^{\prime }=\sum_{x}p(y|x)p_{x}$.
A noiseless channel is such that $\mathcal{X}=\mathcal{Y}$ and
the probabilities $ p(y|x)$ are either 0 or 1 (i.e. it amounts to a permutation
of the alphabet $ \mathcal{X}$).  A particular case is  the
ideal (identity) channel $\mathrm{Id}$, for which $p(y|x)=\delta_{x,y}$ (Kronecker's delta).

\begin{figure}[t]
\begin{center}
\includegraphics[width=330pt]{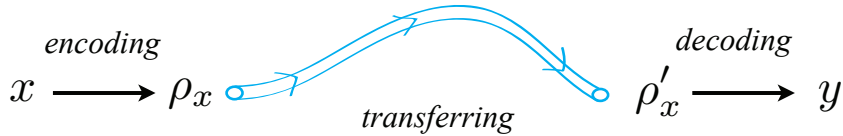}
\caption{Transferring classical data $x\in {\cal X}$ via a quantum information carrier. The first stage of the process requires the encoding of $x$ into a quantum state $\rho_x$ of the
carrier (c-q mapping); then the carrier  propagates along the communication line and its state gets transformed into the density matrix $\rho_x'$ (q-q mapping); finally there is
the decoding stage where the receiver of the message tries to recover $x$ by performing some measurement on $\rho_x'$ and obtaining the classical outcome $y$ (q-c mapping).}
\label{diagramcq}
\end{center}
\end{figure}

Looking for a generalization to the quantum domain we should first consider
those scenarios in which classical data  (i.e. data which could have been
stored into the state of a classical carrier represented by the alphabet ${\cal X}$)
 are transmitted through a
physical communication line which employs  quantum carriers to
convey information (e.g. an optical fiber operating at very low
intensity of the light~\cite{cd,GISIN}). In these configurations the signaling process requires
an initial encoding stage in which the elements of ${\cal X}$ are
``written" into the quantum states of the carrier, and a final
decoding stage in which the received quantum states of the carrier
are mapped back  into classical data by some measurement, as schematically shown on the diagram of Fig.~\ref{diagramcq}.
%%%%%%%%%%%%%%%%%%%%%%%%%%%%%%%%%%%%%%%%%%%%%%%%%%%%
%%%%%%%%%%%%%%%%%%%%%%%%%%%%%%%%%%%%%%%%%%%%%%%%%%
The  encoding stage of such scheme is characterized by
assigning a {\it classical-quantum} (c-q) mapping $ x\rightarrow
\rho _{x}$ that defines which density matrix $\rho_x$ of the quantum
carrier represents the symbol $x \in {\cal X}$ of the input
classical message. Physically, $x$ can thus be interpreted as a parameter of a state
preparation procedure (the correspondence $x\rightarrow \rho _{x}$
containing, in a condensed form, the description of the physical process
generating the state $ \rho _{x}$). Notice that in order to preserve
the statistical structure of the process,  a mixed input state of
${\cal X}$  defined by the probability distribution $P=\left\{
p_{x}\right\} $ is mapped by the c-q channel into the density
operator $ \rho=\sum_{x}p_{x}\rho _{x}$.

On the other hand, the decoding stage
of  Fig.~\ref{diagramcq} is characterized by a
{\it quantum-classical} (q-c) mapping that establishes
the probabilities of the output letters~$y\in{\cal Y}$ corresponding to
the quantum state of the carrier emerging from the quantum
communication line. Such a mapping is implemented by a quantum
measurement and it is characterized by assigning the probability distribution
 $p_y({\rho })$
 that,  given a generic state  $\rho$ of the carrier, defines the statistics of the
possible measurement outcomes.
It can be shown~\cite{ho1} that the linear dependence of
 $p_y({\rho })$
resulting from the preservation of mixtures, along with general  properties of
probabilities, implies the following functional structure
 \begin{eqnarray}
p_y(\rho)=\mathop{\rm Tr}\rho M_{y}\;,  \label{born}
\end{eqnarray}
where $\left\{ M_{y}\right\} $ is a collection of Hermitian
operators in $ \mathcal{H}$ with the properties: $M_{y}\geq
0,\sum_{y}M_{y}=I.$ Any such collection is called probability
operator-valued measure (POVM), or, in the modern quantum
phenomenology, just {\it observable }(with values~$y$)~\cite{h03}. Later we shall explain how the operators $M_{y}$ arise
from the dynamical description of a measurement process. An
observable is called {\it sharp } if $ M_{y}$ are mutually
orthogonal projection operators, i.e. $M_{y}^{2}=M_{y}, M_{y}M_{y'}=0$ if $y\neq y'$. In this case
(\ref{born}) amounts to the well-known ``Born-von Neumann statistical postulate''~\cite{dir58,neu32}.

The c-q and q-c transformations described above are
special cases of quantum channels with classical input, resp.
output, which will be considered in more detail in Sec. \ref{q-c-q}.
The central link in the scheme of Fig.~\ref{diagramcq} is what really
defines the quantum character of the process. It describes the
transformation that the state of the quantum carrier experiences
when propagating through the communication line
 and it is   fully characterized by assigning
a quantum-quantum (q-q) mapping
\begin{eqnarray}
 \label{defphi} \Phi: \;\;\rho \;\; \longrightarrow \;\; \rho'=\Phi \lbrack \rho]\;,
\end{eqnarray}
which takes  input density operators $\rho$ into output density operators $\rho'$
according to the specific physical properties of the model under
consideration, while  respecting the statistical mixtures, i.e.
\begin{equation}\label{aff}
 \Phi{\tiny \left[ \sum_{\alpha}p_{\alpha}\;\rho_{\alpha}\right] }=\sum_{\alpha}p_{\alpha}\; \Phi [\rho_{\alpha}]\;,
\end{equation}
for all probability distributions $\{ p_\alpha\}$ and for arbitrary density operators $\{ \rho_\alpha\}$.
An exact characterization of such transformations is clearly
mandatory as they constitute the proper quantum counterparts  of the
stochastic mapping $P\rightarrow P'$ of the classical communication
theory.
\newline

{\bf Example: } The basic  example of a quantum system is the
{\it qubit} -- a two-level quantum system (say, the spin of an electron) characterized by Hilbert ${\cal H}$ space of dimensionality~$2$.
Via the specification of a canonical basis in ${\cal H}$, its linear operators are represented as
linear combinations of the {\it Pauli matrices}
\begin{eqnarray}
\fl \qquad I\equiv\sigma_0={\small \left[
\begin{array}{cc}
1 & 0 \\
0 & 1
\end{array}
\right],} \; \;\sigma_x={\small \left[
\begin{array}{cc}
0 & 1 \\
1 & 0
\end{array}
\right], }\;\; \sigma_y=
{\small \left[
\begin{array}{cc}
0 & -i \\
i & 0
\end{array}
\right],} \; \;\sigma_z=
{\small \left[
\begin{array}{cc}
1 & 0 \\
0 & -1
\end{array}
\right].}
\end{eqnarray}
This allows us to express the density operators $\rho
\in\mathfrak{S}({\mathcal{H}})$ of the system as points $\vec{a}=(a_x,a_y,a_z)$ in
the unit ball in the three-dimensional real vector space $\mathbb{R}^3$ (the {\it Bloch ball})  by identifying the coordinates $a_x,a_y,a_z$
with the {\it Stokes parameters}  of the expansion (in which the condition $\mbox{Tr} \rho =1$ is taken into account)
\begin{eqnarray}  \label{Sta}
\rho=\frac{1}{2}(I+a_x\sigma_x+a_y\sigma_y+a_z\sigma_z)= \frac{1}{2}
{\small \left[
\begin{array}{cc}
1+a_z & a_x-ia_y \\
a_x+ia_y & 1-a_z
\end{array}
\right] }\;.
\end{eqnarray}
Equation~(\ref{aff}) then implies that a qubit channel is a linear map
in $\mathbb{R}^{3}$ transforming the Bloch ball into certain
ellipsoid inside the ball. Up to rotations at the input and the
output, such a transformation has the following canonical form in
the basis of Pauli matrices:
 \begin{equation}
a_{\gamma}\longrightarrow a'_{\gamma}=b_{\gamma }+a_{\gamma
}t_{\gamma }\;,\qquad\quad \gamma =x,y,z\;,  \label{lam}
\end{equation}
with $b_{\gamma },t_{\gamma }$ real numbers (satisfying some further
constraints to be discussed later). The equation of the output
ellipsoid is
$\sum_{\gamma =x,y,z}\left(\frac{a'_{\gamma}-b_{\gamma }}{t_{\gamma
}}\right)^2=1$.
Hence $b_{\gamma }$ give the coordinates of the center of the output
ellipsoid, while $|t_{\gamma }|$ its half-axes.

\subsection{Dynamics of isolated and open quantum systems}\label{irreversibility}
The simplest case of the q-q mappings (\ref{defphi}) is represented by
 the  {\em noiseless} quantum channels which describe
 reversible transformations of the set of an isolated quantum states  $\mathfrak{S}(\mathcal{H})$  onto itself.
 The famous Wigner's theorem implies~\cite{d2} that any such
 mapping is implemented by either unitary or anti-unitary
conjugation, which amounts~to
\begin{eqnarray}
\Phi \lbrack \rho ]=U\; \rho \; U^\dag\;,
\qquad \mathrm{or\qquad }\Phi \lbrack
\rho ]=U\; \rho ^{\top }\; U^{\dag }\;,
\label{wigner}
\end{eqnarray}
where $U$  is a unitary operator on the system Hilbert space ${\cal
H}$, $(\cdot)^\dag$ denotes
the Hermitian adjoint, while $(\cdot)^{\top }$ is the matrix
transposition in a fixed basis. For a qubit, the mappings in Eq.~(\ref{wigner}) define, respectively,
rotations of the Bloch ball (orthogonal maps with determinant $+1$), and
 reflection with respect to $xz$-plane (i.e. matrix
transposition in the canonical basis) followed by rotations
(orthogonal maps with determinant $-1$).
It turns out however, that only the unitary conjugations
(the first case in~(\ref{wigner})) can be associated with proper
dynamical processes,
the anti-unitary being excluded since
they cannot be continuously connected with the identity mapping (the transposition $\rho \rightarrow \rho ^{\top }$
is indeed typically   identified with time reversal).
\newline

 As it was already mentioned in Sec.~\ref{sec:ClassicalQuantum}, a basic feature of quantum information
as distinct from classical is  the {\it impossibility of
cloning}~\cite{WZ}. Clearly, any classical data can be copied
exactly in  an arbitrary quantity. But a ``quantum xerox'',
i.e. a physical device which would accomplish a similar task for
 arbitrary quantum states contradicts the principles of
Quantum Mechanics. Indeed, the cloning transformation
 \begin{eqnarray} \label{cloning} \ket{\psi}\in {\cal H} \quad \longrightarrow\quad
\underbrace{\ket{\psi}\otimes \dots \otimes \ket{ \psi} }_{n} \in {\cal H}^{\otimes n} \equiv {\cal H}\otimes \cdots \otimes {\cal H} \;,
\end{eqnarray}
is nonlinear and cannot be implemented by a unitary operator
(even if operating jointly on some external ancillary system
that is traced away at the end of the process). Of course,
this can be done for each given state $|\psi\rangle$ (and even for each fixed set of
orthogonal states) by a corresponding specialized device, but there is
no universal cloner for all quantum states. It
should be noted also that approximate cloning of arbitrary quantum states
is allowed, as long as the resulting transformation realizes
Eq.~(\ref{cloning})  to a certain known  degree of accuracy, see
e.g. Ref.~\cite{SIGA}. The quantum xerox is not
the only type of a machine which is forbidden by the laws of Quantum
Mechanics. For a comprehensive list of such ``impossible devices"  we
refer the reader to Ref.~\cite{W}.
\newline

\begin{figure}[t]
\begin{center}
\includegraphics[width=330pt]{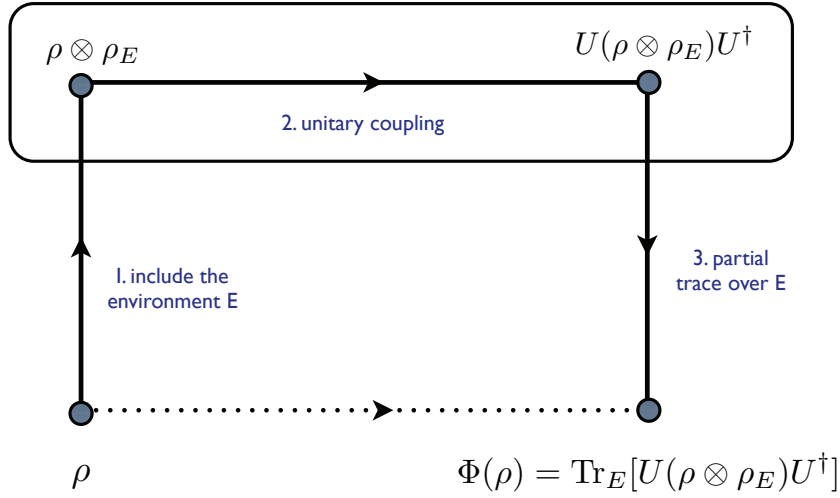}
\caption{Schematics of the unitary representation~(\ref{ivolu}) for the quantum channel $\Phi$.}
\label{figu2}
\end{center}
\end{figure}

The  evolution of an {\it open} system, subject to exterior
influences, whether it be the process of establishing equilibrium
with an environment or interaction with a measuring apparatus,
reveals features of irreversibility. These transformations
constitute the class of {\it noisy} quantum channels~(\ref{defphi})
which cause distortion to the transmitted
states of a quantum carrier in its propagation through the
communication line. A formal characterization of such processes can
be obtained by introducing a Hilbert space $\mathcal{H}_{E}$ to
describe the environmental degrees of freedom $E$ which tamper with
the communication, and assigning to it an initial state $\rho_E$.
Since the carrier and the environment together form an isolated
system, their joint  (reversible) evolution can  be described by a
unitary operator $U$ which, acting nontrivially on the composite
Hilbert space $\mathcal{H}\otimes\mathcal{H}_{E}$, defines their
interaction -- see Fig.~\ref{figu2}. Consequently  the irreversible q-q
mappings~(\ref{defphi})  can be obtained by averaging off $E$ from
the resulting output configurations, i.e.
\begin{eqnarray}
\rho\; \longrightarrow\;  \Phi [\rho ]={\rm Tr}_{E}\left[ U\left(
\rho \otimes \rho _{E}\right) U^{\dag } \right] \;\label{ivolu}.
\end{eqnarray}
%%%%%%%%%%%%%%%%%%%%%%%%%%%%%%%%%%%%%%%%%%%%%%%%%%%%
%%%%%%%%%%%%%%%%%%%%%%%%%%%%%%%%%%%%%%%%%%%%%%%%%%
This expression   provides what is generally called a {\em unitary
representation} for the quantum channel $\Phi$ and generalizes  the
unitary evolutions of  Eq.~(\ref{wigner})  to the case of
irreversible dynamics. Let us then assume that the initial
environment state is pure,  i.e. $\rho _{E}=|\psi _{E}\rangle \langle \psi _{E}|$ -- via purification of
$\rho_E$ this is always possible: the resulting expression is often
called the Stinespring representation of $\Phi$ \cite{stine,kr}.
In this case Eq.~(\ref{ivolu}) can be written as
\begin{eqnarray}
\Phi [\rho ]={\rm Tr}_{E} V \rho V^{\dag}\;, \label{isometric}
\end{eqnarray}
 with $V$ being  an  isometric operator from ${\cal H}$ to ${\cal H}\otimes {\cal H}_E$ defined by
 \begin{eqnarray} \label{defViso}
 V \equiv U|\psi_E\rangle= \sum_{k=1}^{d_E} V_k \otimes |e_E^k\rangle\;,
  \end{eqnarray}
where  $d_E$ is the dimensionality of ${\cal H}_E$ while  $\left\{ |e^{k}_E\rangle\right\}_{k=1,\cdots, d_E}$ is one of its  orthonormal basis
(recall that a linear operator from one space to another
is isometric if it preserves the norms and hence inner
products of vectors. In particular it satisfies the relation $V{^\dag} V=I$ with $I$ being the identity operator of the {\it input} space).

In this expression   $V_{k}=\langle e^{k}_E|U|\psi _{E}\rangle$ are operators
acting on $\mathcal{H}$ which are uniquely determined by the
relation
\begin{eqnarray}
\langle \varphi |V_{k}|\psi \rangle =\langle \varphi \otimes
e^{k}_E|U|\psi \otimes \psi _{E}\rangle\;, \quad |\varphi\rangle ,|\psi\rangle \in \mathcal{H
}. \label{definitionofVk}
\end{eqnarray}
They are called Kraus operators and satisfy the completeness relation
 \begin{eqnarray} \label{completness}
 \sum_{k=1}^{d_{E}}V_{k}^{\dag}V_{k}=I \;.
 \end{eqnarray}
Taking the partial trace in (\ref{isometric}) in the basis
 $\left\{ |e^{k}_E\rangle\right\}_{k=1,\cdots, d_E}$  we finally arrive at the so called
  {\it operator-sum} (or {\it Kraus})
{\it decomposition}~\cite{kr} for the evolution~(\ref{ivolu})
 \begin{eqnarray}
\Phi \lbrack \rho ]=\sum_{k=1}^{d_{E}}V_{k}\rho V_{k}^{\dag}\;.
\label{kraus}
\end{eqnarray}

It is important to notice that
given a quantum channel $\Phi$ expressed as in Eq.~(\ref{kraus}) with
some set of operators  $\{ V_k\}$, it is always possible to extend it (in a highly non-unique way)
to the open system dynamics~(\ref{ivolu}) with some
effective environment $E$ initialized in a state $\rho_E$, and a
unitary transformation $U$ that couples it to the system (or equivalently  in terms of an isometric transformation $V$ that maps  ${\cal H}$
into ${\cal H}\otimes {\cal H}_E$ as in Eq.~(\ref{isometric})).
Furthermore, even the Kraus decomposition for a given channel $\Phi$ is not unique. Indeed given
a Kraus set $\{V_k\}$ for $\Phi$, a new Kraus set $\{ W_j\}$, which
represents the same quantum channel,   can be obtained by forming
the following linear combinations
\begin{eqnarray}
W_{j}=\sum_{k=1}^{d_{E}}u_{jk}V_{k}\;, \label{Vunitk}
\end{eqnarray}
where $\left[ u_{jk}\right] $ is any complex matrix which satisfies
the isometry constraint \mbox{$\sum_{j} \bar{u}_{k'j} u_{jk} =
\delta_{kk'}$}.
  Let us stress that here, as anywhere in quantum information science,
``environment'' means those degrees of freedom of the actual
physical environment of the open quantum system which are
essentially involved in the interaction and in the resulting
information exchange with the system (cf. the notion of ``faked
continuum'' in~\cite{Sten}). Therefore the question of such a
{\it minimal environment} for a given channel arises quite
naturally (clearly the Kraus decomposition with the minimal number
of non-zero components $d_{E}$ is unique only up to transformations
of the form~(\ref{Vunitk}) with $u_{jk}$ being now unitary matrices;
on the contrary, the size $d_{E}$ of the minimal environment is an
important invariant of the channel $\Phi$ reflecting its
``noisyness'' and irreversibility). In the minimal decomposition the
Kraus operators are linearly independent, so that their number
cannot exceed the dimensionality $d^2$ of  the space
of linear operators acting on the system -- for instance, the
reversible channels of Eq.~(\ref{wigner}) have a minimal
decomposition with a single Kraus element (the unitary $U$ which
defines them).
\newline 

{\bf Example 1:} The {\it depolarizing} channel (with probability of error $p$) is
given by the formula (in which $\rho$ is an arbitrary operator, not necessarily of unit trace)
\begin{eqnarray}
\Phi \lbrack \rho ]=(1-p)\rho +p
\frac{I}{d}\; \mathop{\rm Tr}\rho \;,  \label{dep}
\end{eqnarray}
where $\dim {\mathcal{H}}=d$. This relation describes mixture of the
 ideal channel $\mathrm{Id}: \rho \rightarrow \rho$ and of the completely depolarizing
channel $\rho\rightarrow (I/d) \mathop{\rm Tr}\rho$
which transforms any state $\rho$ into the chaotic
state~${I}/{d}$. For the depolarizing channel a Kraus decomposition
can be obtained by writing the unit operator and the trace in a
fixed  basis $\{ |e^j\rangle\}$:
 \begin{eqnarray}
\Phi \lbrack \rho ]
=(1-p)V_{0}\rho
V_{0}^{\dag }+\frac{p}{d}\sum_{i,j=1}^{d}V_{ij}\rho V_{ij}^{\dag }\;,
\label{depkraus}
\end{eqnarray}
where $V_{0}=I,V_{ij}=|e^{i}\rangle \langle e^j|.$
Such Kraus set is clearly not minimal
since its elements are linearly dependent (e.g. $V_{0}=\sum_{i=1}^{d}V_{ii}$).
\newline 

{\bf Example 2:} As another example consider the qubit channel (\ref{lam}) with
$b_{\gamma }\equiv 0,$ that contracts the Bloch ball along the axes
$\gamma=x,y,z$ with the coefficients $|t_{\gamma}|$ (notice that
qubit depolarizing channel corresponds to the uniform contraction
with $t_x=t_y=t_z=1-p$). By using the multiplication rules for Pauli
matrices one sees that in this case
\begin{eqnarray}
\Phi \lbrack \rho ]=\sum_{\gamma =0,x,y,z}p_{\gamma }\; \sigma
_{\gamma }\; \rho\;  \sigma _{\gamma },  \label{dia}
\end{eqnarray}
where
\begin{eqnarray}
&&p_{0}=\left( 1+t_{x}+t_{y}+t_{z}\right)/4 \;,\qquad
p_{x}=\left( 1+t_{x}-t_{y}-t_{z}\right)/4 \;, \\
&&p_{y}=\left( 1-t_{x}+t_{y}-t_{z}\right)/4 \;,\qquad
p_{z}=\left( 1-t_{x}-t_{y}+t_{z}\right)/4 \;,
\end{eqnarray}
and the nonnegativity of these numbers is the necessary and sufficient
condition for the complete positivity of the map $\Phi$ (see below). In this
case (\ref{dia}) gives the minimal Kraus decomposition for the channel  (the size of the minimal environment being equal to the number of strictly positive coefficients $p_{\gamma}$).
\newline
\newline

Coming back to the
decomposition (\ref{kraus}) we see that
it can be written as
\begin{eqnarray}
\Phi \lbrack \rho ]
=\sum_{k=1}^{d_{E}}p_{k}(\rho) \; \rho _{k}\;,  \label{decom}
\end{eqnarray}
where $p_k  (\rho)={\rm Tr}\rho M_{k}$
is the probability distribution associated with a POVM of elements $M_k= V_k^\dag V_k$, and $\rho _{k} = V_{k}\rho V_{k}^{\dag}/p_k  (\rho)$  are density operators.
Noticing  that  $V_{k}\rho V_{k}^{\dag}=\langle e_{E}^k|U\left( \rho \otimes |\psi_{E}\rangle \langle \psi _{E}|\right) U^{\dag }|e_{E}^k\rangle$,
the relation (\ref{decom}) can then be interpreted as follows: after the system
and the environment have evolved into the state
$U\left( \rho \otimes |\psi_{E}\rangle \langle \psi _{E}|\right) U^{\dag}$ via the unitary coupling $U$,
in the environment a von Neumann measurement in the basis $\left\{
|e^{k}_{E}\rangle \right\}$ is performed \cite{neu32}; the outcome~$k$ appears
with probability  $p_k  (\rho)$ and the {\it posterior state} of the
system conditioned upon the outcome $k$ is the density operator
$\rho _{k}.$ In other words the formula (\ref{decom}) gives the decomposition
of the system state $\Phi \lbrack \rho ]$ after the (nonselective)
measurement into ensemble of posterior states corresponding to
different measurement outcomes.

\subsection{Heisenberg picture} \label{heisenbergp}
 So far we worked in the Schr\"odinger picture describing
evolutions of states for fixed observables. The passage to the Heisenberg
picture where the observables of the system evolve while the states are kept fixed, is obtained by introducing the dual channel
 $\Phi^*$ according to the rule
 \begin{eqnarray}
 {\rm Tr} \Phi \lbrack \rho ]\; X = {\rm Tr} \rho \;\Phi^*[X] \;, \label{Vheisenberg}
\end{eqnarray}
which must hold for all density operators $\rho\in  \mathfrak{S} ({\cal H})$ and for all operators $X$ in ${\cal H}$.
The map $\Phi^*$ is also completely positive admitting the operator-sum decomposition
\begin{eqnarray}
\Phi^*[X]=\sum_{k}V_{k}^\dag X V_{k}\;,\label{VKRAUSHE}
\end{eqnarray}
where   $\{V_k\}$ is a Kraus set for $\Phi$. The property of trace preservation for $\Phi$  is equivalent to the fact that $\Phi^{\ast}$
is {\it unital}, i.e. it leaves invariant the unit operator $I$: $\Phi^*[I]=I$.
A channel $\Phi$ in the Schr\"odinger picture may also occasionally
be unital, in which case it is called {\it bistochastic}; it
leaves invariant the chaotic state $I/d$. This happens
always when the channel is {\it self-dual}: $\Phi= \Phi^*$.
An example of such a channel is provided by the depolarizing map of
Eq.~(\ref{dep}).

\subsection{Positivity, complete positivity  and the Choi-Jamiolkowski representation}

In accordance with the fact that the channel $\Phi$ must  transform
quantum states into quantum states,
Eq.~(\ref{kraus}) ensures preservation of positivity  (indeed $\rho
\geq 0$ implies $V_{k}\rho V_{k}^\dag \geq 0$), while  Eq.~(\ref{completness})
guarantees that the mapping $\Phi$ preserves the trace
(i.e. $\mathop{\rm Tr}\Phi [ \rho]=\mathop{\rm Tr}\rho$). Thus, for
arbitrary linear operators $\rho$ in ${\cal H}$, the transformation
(\ref{kraus}) defines a positive, trace preserving, linear mapping.

Quite importantly, it satisfies an even stronger property called
{\it complete positivity } which means that any extension $\Phi
\otimes\mathrm{Id}_{R}$ of the channel $\Phi $ by an ideal channel
$\mathrm{Id}_{R}$ of a ``parallel'' reference system $R$ is  again
positive. In other words, given a quantum channel $\Phi$ operating
on the system $A$ which admits the Kraus decomposition, for any
state $\rho _{AR}$ of the composite system $\mathcal{H}_A\otimes
\mathcal{H}_{R}$ the equation
\begin{eqnarray}
(\Phi \otimes \mathrm{Id}_{R})[\rho _{AR}]=\rho_{AR}^\prime  \label{etrans}\;,
\end{eqnarray}
 defines again a proper density operator $\rho _{AR}^\prime$ (indeed, $(\Phi \otimes
\mathrm{Id}_{R})[\rho _{AR}]=\sum_{k=1}^{d_{E}}(V_{k}\otimes
I_{R})\rho _{AR}(V_{k}\otimes I_{R})^\dag \geq 0$ while the trace is
obviously preserved). The complete positivity together with the
trace preservation  (CPTP in brief) are {\it necessary and
sufficient} for a linear map  $\Phi$ to have a Kraus
decomposition~(\ref{kraus})  and thus, a unitary
representation~(\ref{ivolu}); hence the CPTP conditions are
characteristic for the evolution of an open system and can be used
to define a quantum channel.

When the initial state $\rho_{AR}$ is nonseparable, the
transformation described in Eq.~(\ref{etrans}) is called
``entanglement transmission''. At the physical level it represents
those processes by which the sender creates locally
some entangled configuration of the systems $A$ and $R$, and then
sends half of it (specifically the carrier $A$) to the receiver
through a q-q channel. The fact that the mapping $\Phi$ entering
in Eq.~(\ref{etrans}) is completely positive  guarantees that the
resulting configuration can still be represented as a density
operator, and hence that the overall transformation admits a quantum
mechanical description. Notably not all positive maps which take the
set of states $\mathfrak{S} ({\cal H}_A)$ of a given system $A$ into
itself are completely positive. The canonical example is the
transposition $T[\rho_A]=\rho_A^{\top}$ in a fixed basis
 $\{ |e_j^A\rangle\}$, see  the second relation in Eq.~(\ref{wigner}). Since $\rho_A^{\top}$ and
$\rho_A$ share the same spectral properties, it is clear that $T$
maps the states of $A$ into states  (hence it is positive). However
applying the map $T \otimes \mathrm{Id}_{R}$ to a maximally
entangled state  $\frac{1}{\sqrt{d_A}}\sum_{j}\ket{e^j_A\otimes
{e^j_R}}$ of the composite system $AR$ one obtains the operator
\begin{eqnarray} \nonumber
\frac{1}{d_A}(T\otimes \mathrm{Id}_{R})\left[ \sum_{i,j=1}^{d_A}|e^{i}_A
\otimes e^{i}_R\rangle \langle e^{j}_A\otimes  e^{j}_R|
\right] =\frac{1}{d_A}\sum_{i,j=1}^{d_A}|e^{j}_A\rangle \langle
e^{i}_A|\otimes |e^{i}_R\rangle \langle e^{j}_R| \;,
\end{eqnarray}
which is not a proper state of $AR$ since it is not positive
(its expectation value on the vector $|e_A^1\otimes e_R^2\rangle -  |e_A^2\otimes e_R^1\rangle$ is negative).
Such a weird behavior of $T$ should not be
surprising: as anticipated at the beginning of the section,  in
physics transposition is related to the time inversion. The operation
$T\otimes \mathrm{Id}_{R}$ is thus something like doing time
inversion only in one part of the composite system $AR$.
To avoid this
unphysical behavior, one should hence work under the explicit
assumption that the q-q mappings entering the Eq.~(\ref{defphi})
satisfy  to the complete positivity condition.

In the case of CPTP mappings, the right hand side of
Eq.~(\ref{etrans}) obtained by substituting the maximally entangled
state associated with the vector
$\frac{1}{\sqrt{d_A}}\sum_{j}\ket{e^j_A\otimes {e^j_R}}$ for $\rho_{AR}$, provides  the {\it Choi-Jamiolkowski state} of~$\Phi$:
 \begin{eqnarray} \label{choi}
\rho^{(\Phi)}_{AR}=\frac{1}{d_A}\sum_{i,j=1}^{d_A} \Phi \left[ |e^{i}_A\rangle
\langle e^{j}_A|\right] \otimes |e^{i}_R\rangle \langle e^{j}_R|\;,
\end{eqnarray}
which uniquely determines the channel through the inversion formula
\begin{eqnarray}
\Phi \lbrack {\rho_A}]=d_A\;  {\rm Tr}_{R}\left[ \rho^{(\Phi)}_{AR}(I_{A}\otimes
{\rho}_{R}^{\top })\right]\;.
\end{eqnarray}
Here ${\rho}_{R}^{\top }=\sum_{j,k}\braket{e^k_A}{{\rho}_A|e^j_A}
\ketbra{e^j_R}{e^k_R}$ is the state of $R$ which is constructed via
transposition of the density matrix of $\rho_A$.

\subsection{Compositions rules}

Suppose that we have a family $\{\Phi_{\alpha}\}$ of quantum channels
where the values of the parameter $\alpha$ appear with probabilities
$p_{\alpha}$. Then the mapping defined as
\begin{eqnarray}
\Phi[\rho] = \sum_{\alpha} p_{\alpha}\;  \Phi_{\alpha}[\rho]\;,
\end{eqnarray}
is still CPTP and hence defines a channel (with a fluctuating
parameter $\alpha$). In other words, the set of channels is convex,
i.e. closed under convex combinations (mixtures). The extremal
(pure) channels are those CPTP transformations which admit a Kraus
decomposition~(\ref{kraus}) with $V_k$ such that $\{ V_k
V_{k^{\prime}}^\dag\}_{k,k'}$ form a collection of  linearly
independent operators \cite{choi} (in particular, the unitary
evolutions~(\ref{wigner}) are extremal).

The set of channels possesses also a semigroup structure under
concatenation. Indeed the transformation $\Phi_2 \circ
\Phi_1$ obtained by applying $\Phi_2$ at the output of the quantum
channel $\Phi_1$, i.e.
\begin{eqnarray}
(\Phi_2 \circ \Phi_1)[\rho]  = \Phi_2 \left[ \Phi_1[\rho]  \right] \label{comp}\;,
\end{eqnarray}
is still CPTP and hence a channel. Notice that the product
``$\circ$" is in general  neither commutative (i.e.  $\Phi_2 \circ
\Phi_1 \neq \Phi_1\circ \Phi_2$), nor it admits inversion (the only
quantum channels  which admit CPTP inverse are the reversible
transformations defined by the unitary mappings~(\ref{wigner})). The
possibility of concatenating quantum channels with unitary
transformations turn out to be useful in defining equivalence
classes which, in several cases, may help in simplifying the study
of the noise effects acting on a system. In particular two maps
$\Phi_{1}$ and $\Phi_2$ are said to be {\it unitary equivalent} when
there exists unitary transformations $V$ and $U$ such that $\Phi_2 =
\Phi_U\circ \Phi_1 \circ \Phi_V$ (here $\Phi_U[\cdot] \equiv U
[\cdot] U^\dag$ is the quantum channel associated with the unitary
$U$, similarly for $\Phi_V$). For instance, exploiting this
approach,  classifications for the quantum channels operating on a
qubit~\cite{rsw} or on a single Bosonic
mode~\cite{cg,hclass,cgh} have  been realized (see also
Sec.~\ref{onemode}).

Finally,  the complete positivity allows us to consider tensor
products of channels which are most important in information theory,
where tensor products describe parallel or block channels. Indeed,
one can write the tensor product as concatenation of completely
positive maps
 \begin{eqnarray} \label{compo1}
 \Phi _{1}\otimes \Phi _{2}=\left( \Phi _{1}\otimes \mathrm{Id}_{2}\right)
\circ \left( \mathrm{Id}_{1}\otimes \Phi _{2}\right) \;,
\end{eqnarray}
which is again completely positive.

\subsection{Complementary channels} \label{complement}

 For simplicity till now we considered channels with the same
input and output spaces. From the point of view of information
theory, however, it is quite natural to have possibility
for them to be different. This is done by considering
transformations~(\ref{kraus}) with operators $V_{k}$ mapping the
input space $\mathcal{H}_{A}$ into the output space
$\mathcal{H}_{B}$ while still satisfying the completeness relation~(\ref{completness}), or equivalently,  by considering in~(\ref{isometric}) isometric transformation $V$ which connects  ${\cal H}_A$ to
${\cal H}_B \otimes {\cal H}_E$. In such cases we shall often use self-explanatory notation $\Phi= \Phi_{A\rightarrow B}$.
\newline

{\bf Example:} The {\it quantum erasure
channel}~\cite{grassl,erasure} gives a natural example of a channel with
different input and output spaces. It represents a
communication line which transmits the input state $\rho$ intact
with probability $1-p$ and ``erases''  it with probability $p$ by
replacing it with  an erasure signal $|e\rangle$ that is orthogonal
to $\rho$. Formally it can be defined  as the CPTP map
\begin{eqnarray} \label{erasure}
\rho\quad \longrightarrow \quad  (1-p)\; \rho \oplus p\;  |e\rangle\langle e| \; \mbox{Tr}\rho \;,
\end{eqnarray}
which operates  from the input space ${\cal H}_A$ to  the output space
${\cal H}_B = {\cal H}_A \oplus \{ |e\rangle\}$ obtained by ``adding'' an extra orthogonal vector $|e\rangle$ to
${\cal H}_A$.
\newline

A quite important case where the input and output systems differ arises
 in the open system description (\ref{ivolu})  of a channel $\Phi$ that maps a system $A$  into itself.
Indeed  suppose that, instead of the  evolution of $A$, we are interested in the  state change of the
environment $E$ as a function of the input state $\rho$ of $A$.  This is given by the following mapping
\begin{equation}\label{VCOMP1}
\rho\quad \longrightarrow \quad \mathrm{Tr}_{A}U(\rho \otimes \rho _{E})U^{\dag },
\label{fie}
\end{equation}
where in contrast to Eq.~(\ref{ivolu}), the partial trace is taken with respect to $\mathcal{H}=\mathcal{H}_A$ instead of $\mathcal{H}_{E}$.
The transformation (\ref{fie}) defines a quantum channel that connects states in ${\cal H}_A$ to states of the channel environment.
There is a complementarity relationship  between the mappings (\ref{ivolu})  and (\ref{fie})
which reflects various aspects of
information-disturbance trade-off in the evolution of open quantum
system. This relation becomes especially strict in the case of pure environmental
state $\rho _{E}$ when it amounts to the notion of {\it complementary}
channels (if  $\rho_E$ is a mixed state such complementarity is weakened and   the two maps are sometimes called {\it weakly
complementary} \cite{cgh}).
More generally, complementary channels can be defined also in a three quantum systems  settings. Given  the spaces $\mathcal{H}_{A},
\mathcal{H}_{B},\mathcal{H}_{E}$ that represent the systems $A$, $B$, $E$,  and  an isometric operator $V:\mathcal{H}
_{A}\rightarrow \mathcal{H}_{B}\otimes \mathcal{H}_{E}$ which connects them, we say that the mappings
 \begin{eqnarray}
\Phi(\rho )=\mathrm{Tr}_{E}V\rho V^{\dag }\;,\qquad \qquad \qquad
\tilde{\Phi}(\rho )=
\mathrm{Tr}_{B}V\rho V^{\dag }\;,\quad  \label{compl}
\end{eqnarray}
with  $\rho$ being a density operator in $\mathcal{H}_{A},$ define two channels which
are {\it complementary}
\cite{ds,h,mr} (notice that they reduce to Eqs.~(\ref{ivolu}) and (\ref{VCOMP1}),
respectively,  in the case $B=A$).
Expressing the isometry as in  Eq.~(\ref{defViso})
 where again $\{ |e_E^k\rangle\}_k$  is an orthonormal basis of ${\cal H}_E$ while  the $V_k$'s are now  operators from ${\cal H}_A$ to ${\cal H}_B$,
the above relations can be rewritten as
 \begin{eqnarray} \label{cmpm}
\Phi[\rho ]=\sum_{k =1}^{d_E}V_{k}\rho V_{k }^{\dag }\;, \qquad \quad
{\tilde{\Phi}}[\rho ]=\sum_{k ,\ell =1}^{d_E}
\;|e_{E}^k\rangle \langle
e_{E}^\ell| \;  \mathrm{Tr} \left[
\rho V_{\ell}^{\dag }V_{k }\right]  \;,
\label{kr}
\end{eqnarray}
the first one being a Kraus decomposition for $\Phi$.
It can be shown that the second relation in Eq.~(\ref{kr}) determines all  complementary channels of $\Phi$ uniquely up to
unitary equivalence, moreover, by using a similar construction for
$\tilde{\Phi}$ one can check that its complementary is
 $\tilde{\tilde{\Phi}}=\Phi$. Similar relation does not hold for the case of  weakly
complementary channels which are not unique and depend on the selected unitary
representation  (\ref{ivolu}) of the original map $\Phi$.
\newline

{\bf Example:}  A complementary channel associated with the identity map Id on ${\cal H}$
(i.e. $\mbox{Id}[\rho]= \rho$ for all $\rho$),
is any map $\tilde{\Phi}$ that transforms all $\rho$  into a fixed pure output state $|\psi_E\rangle$ of the environment, i.e.
\begin{eqnarray}
\tilde{\Phi}(\rho )
 = |\psi_E\rangle \langle \psi_E| \;
\mathrm{Tr}\rho \;  , \label{complid}
\label{cid}
\end{eqnarray}
(to see this put ${\cal H}_A={\cal H}_B$ in Eq.~(\ref{compl}) and take
the isometry $V: {\cal H}_A \rightarrow {\cal H}_A\otimes {\cal H}_E$ to be the operator
$V= I \otimes |\psi_E\rangle$). Further, a channel that transforms the density operators $\rho$
into a fixed non-pure output state $\rho_E$ in ${\cal H}_E$ is weakly complementary
to the identity channel $\mathrm{Id}$. (Similarly one finds that complementary to the
erasure channel~(\ref{erasure})  is  again an  erasure channel where parameter $p$ has been replaced by $1-p$).

This example illustrates the fact that perfect transmission of
quantum information through a quantum channel is equivalent to absence
of quantum information transfer from the input to the environment $E$, and vice
versa. An approximate version of this complementarity expresses the quantum principle
of {\it information-disturbance trade-off}: the closer is the channel $\Phi
$ to an ideal channel, the closer its complement $\tilde{\Phi}$ to
the completely depolarizing one \cite{dar}.

\subsection{Quantum channels involving a classical stage}
\label{q-c-q}

As shown in Fig.~\ref{diagramcq}, the process of transmission of classical information through a quantum  communication line
includes encoding and decoding stages which are special cases of quantum channels
with classical input, resp. output. Furthermore, a quantum information
processing system such as a quantum computer, may include certain sub-routines
where the data have essentially classical nature. Here we characterize such mappings by describing them as special instances of CPTP maps.
\begin{itemize}
\item[(i)] The classical-quantum (c-q) channel. In Sec.~\ref{Sec22} we have seen that this class of
channels is uniquely defined by a map $x\rightarrow \rho _{x}$
describing encoding of the classical input $x\in{\cal X}$ into quantum state
$\rho _{x}$ in the output space $\mathcal{H}_B$.  This  process can
be represented  as a quantum channel by introducing the
input space $\mathcal{H}_A$ spanned by an orthonormal basis
$\{\ket{e_{A}^x}\}_x,$ so that
 \begin{eqnarray}\label{defCQ}
\rho \qquad \longrightarrow \qquad \Phi_{A\rightarrow B}^{\mbox{\tiny{(c-q)}}}
 \lbrack \rho ]=\sum_{x\in{\cal X}}
\braket{e_A^x}{\rho\vert e_A^x}\rho _{x}\;, \end{eqnarray}
where $\rho $ is a density
operator in $\mathcal{H}_A$ (accordingly output ensembles of the form $\rho=\sum_x p_x \rho_x$ are generated  by taking the input states $\rho$ such that $\braket{e_A^x}{\rho\vert e_A^x}=p_x$).
An extreme case of c-q channel is  the completely depolarizing channel,
describing the ultimate result of an irreversible evolution to a
final fixed state as in Eq.~(\ref{complid}).

\item[(ii)] The quantum-classical (q-c) channel. Similarly to the previous case, (q-c) transformations can be represented as a CPTP map  connecting an input quantum system $B'$ to an output
quantum system $A'$, by introducing for the latter
an orthonormal basis $\{\ket{e_{A'}^y}\}_{y\in{\cal Y}}$ to represent the  classical outcomes ${\cal Y}$,
\begin{eqnarray}
\rho \qquad \longrightarrow \qquad \Phi_{B'\rightarrow A'}^{\mbox{\tiny{(q-c)}}}  \lbrack \rho ]=\sum_{y\in {\cal Y}}\ketbra{e_{A'}^y}{e_{A'}^y}\;  {\rm Tr}\rho M_{y}\;,
\end{eqnarray}\label{meas}
where $\{M_y\}$ is a POVM in the input space  $\mathcal{H}_{A'}$.
\end{itemize}

The transmission of classical information through a quantum communication line can now be modeled as  concatenation of
such maps, i.e. $\Phi_{B'\rightarrow A'}^{\mbox{\tiny{(q-c)}}}   \circ \Phi_{B\rightarrow B'} \circ \Phi_{A\rightarrow B}^{\mbox{\tiny{(c-q)}}}$
with $\Phi_{B\rightarrow B'}$ being a generic CPTP map from $B$ to $B'$.
Of particular interest in quantum information theory is also the class of maps that arise when  the ordering of such a concatenation is reversed. In particular,
identifying $A$ with $A'$ and
assuming that the bases  $\{\ket{e_{A}}^x\}_x$ and $\{\ket{e_{A}}^y\}_y$ coincide,
we obtain a q-c-q channel from $B'$ to $B$ of the form
\begin{eqnarray}
\rho \qquad \longrightarrow \qquad \Phi_{B'\rightarrow B}^{\mbox{\tiny{(q-c-q)}}}  [\rho] =
\left( \Phi_{A\rightarrow B}^{\mbox{\tiny{(c-q)}}} \circ \Phi_{B'\rightarrow A}^{\mbox{\tiny{(q-c)}}} \right) [\rho] =
\sum_{x} \rho _{x} \; {\rm Tr}\rho M_{x}\;,
\label{ebr}
\end{eqnarray}
which describes quantum measurement on $B'$ followed by a state preparation on $B$ depending on the
outcome of the measurement. Such mappings include the c-q and q-c channels as special cases (i.e. c-q channels are obtained by
taking  $M_x$ to be projections onto the vectors of orthonormal basis, while q-c channels are obtained when $\rho_x$ are orthogonal pure states).

The channels defined above are characterized by the property of {\it entanglement-breaking}: operating with them on the half of entangled state $\rho_{B'R}$ of a composite system $B'R$
produces unentangled state of $BR$  (in particular, their Choi-Jamiolkowski state~(\ref{choi}) is separable).
This is not surprising, because in the q-c-q channel the quantum information passes through an intermediate stage (represented by the system $A$) in which information can only be encoded as a classical state.
Less obvious is the fact that any entanglement-breaking channel has the form (\ref{ebr})~\cite{HRS}.
Quite non-obviously, the depolarizing channel (\ref{dep}) is
entanglement-breaking if the error probability $p\geq d/(d+1)$~\cite{bsst}.
\newline

In all the cases described above the complete positivity can be seen by explicitly producing a
Kraus decomposition for the map. For instance, let $\Phi $ be an entanglement-breaking channel
of the form~(\ref{ebr}). Then  by making the spectral decompositions of the
operators $\rho_{x}, M_{x}$ one arrives to a Kraus decomposition
with rank one operators of form
\begin{eqnarray}
 \Phi[\rho] =\sum_{\alpha}|\varphi^{\alpha}\rangle \langle \psi^{\alpha}
|\rho |\psi^{\alpha}\rangle \langle \varphi^{\alpha}|\;,
\label{eb}
\end{eqnarray}
where $|\varphi^{\alpha}\rangle$ are unit vectors while $|\psi^{\alpha}\rangle$
form an {\it overcomplete system} for the  input space,~i.e.
\begin{eqnarray}\label{overcomp}
\sum_{\alpha}|\psi^{\alpha}\rangle  \langle \psi^{\alpha}|=I\;.
\end{eqnarray}
According to the general formula (\ref{cmpm}), the complementary map associated with Eq.~(\ref{eb}) is now obtained as
\begin{eqnarray}
\tilde{\Phi}[\rho]
=  \sum_{\alpha ,\beta}c_{\alpha \beta }|e^{\alpha
}\rangle \langle \psi^{\alpha}|\rho |\psi^{\beta}\rangle \langle
e^{\beta}|\;,  \label{gdiag}
\end{eqnarray}
where $\{\ket{e^\alpha}\}$ is an orthonormal set, and $c_{\alpha \beta }=\langle \varphi^{\beta }|\varphi^{\alpha
}\rangle $ is a nonnegative semi-definite matrix with units on the diagonal.
In the case where the input and output spaces of $\tilde{\Phi}$ coincide and  $\ket{\psi^{\alpha
}}=\ket{e^{\alpha}}$, such channels amount to
elementwise multiplication of the matrices $[\langle
e^{\alpha }|\rho |e^{\beta}\rangle ]$ and $[c_{\alpha \beta }]$, called Schur (or Hadamard) product. Such
channels are called {\it dephasing} \cite{ds} (or {\it diagonal}
\cite{mr}) since they suppress the off-diagonal elements of the
input density matrix. From (\ref{eb}) we see that the dephasing channels are
complementary to a particular class of entanglement-breaking
channels, namely to c-q channels. For another special subclass of
the q-c channels, $\left\{ \ket{\varphi^{\alpha}}\right\}$ of Eq.~(\ref{eb}) constitute an
orthonormal base, so that $c_{\alpha \beta
}=\delta _{\alpha \beta }$ and the complementary channel is a
q-c channel which is called {\it completely dephasing} since it amounts to nullifying the off-diagonal elements
of the density matrix $[\langle e^{\alpha }|\rho |e^{\beta}\rangle]$.

\section{Bosonic Gaussian channels}\label{gaussianchannels}

Until now we have considered quantum channels  from rather abstract point of view
focusing on  models in which the quantum carriers are effectively described as finite dimensional systems (say, collections of qubits).
In many real experimental scenarios such a description is valid only approximately.
To begin with, the fundamental physical information carrier is the electromagnetic field which
is known to be mathematically equivalent to an ensemble of oscillators that in Quantum Mechanics are described
 as infinite dimensional systems~\cite{gla63,kla68,loui,YAMA}.
This is a typical example of ``continuous variables'' Bosonic quantum system~\cite{CONT,cerf} whose basic observables
(oscillator amplitudes) satisfy the Canonical Commutation Relations (CCR) (other examples include vibrational modes of solids, atomic ensembles,
nuclear spins in a quantum dot and Bose-Einstein condensates). Many of the
current experimental realizations of quantum information processing are
carried out in such systems~\cite{cd,CONT,cerf,GISINREV,GISIN}. In particular, the most impressive
examples of quantum communication channels have been realized in this way by using
optical fibers~\cite{TITTEL,MARCI,TAKE,URSIN1}  or free space communication~\cite{SCHMITT,URSIN,PENG}.

In the context of quantum optical communication~\cite{cd} particularly important are the {\it Gaussian states}, including
coherent and squeezed states realized in lasers and nonlinear
quantum optical devices, and the corresponding class of quantum
information processors -- the {\it Gaussian Channels}~\cite{hw,ew,cegh,WANG,WEEDBROOK}.
These last provide the proper mathematical description for the most common sources
of noise  encountered  in optical implementations, including attenuation, amplification
and thermalization processes~\cite{hw,cd}. This Section is devoted to review
of the basics properties of this special class of channels.

\subsection{Example: channel with additive Gaussian quantum noise}
\label{onemodeexa}

As a starter consider the case of a single electromagnetic  mode which describes the
propagation of photons of a given frequency $\omega$ and fixed polarization through
an optical fiber. The physics of the system is fully determined by the annihilation and creation operators
\begin{eqnarray}
a=\frac{1}{\sqrt{2 \omega }}\left( \omega Q+iP\right) \;, \qquad \quad a^{\dagger
}=\frac{1}{\sqrt{2 \omega }}\left( \omega Q-iP\right)\; ,
\end{eqnarray}
where $Q,P$ are the canonical operators (quadratures) which effectively represent the field and satisfy the Heisenberg CCR
\begin{eqnarray}
[ Q,P]=iI \;,  \label{heiccr}
\end{eqnarray}
 (in the Schr\"odinger representation $Q=x$ and $P=i^{-1}d/dx$, so that the physical momentum operator is $p=\hbar P$).
 Suppose that initially the mode is in the thermal (Gibbs) state represented by the density matrix
\begin{eqnarray}
\rho _{0}=\frac{\exp \left[ -\beta H\right] }{\mathrm{Tr}\exp \left[
-\beta H\right] }\;,  \label{qupo}
\end{eqnarray}
where $H=\frac{1}{2 }\left( \omega ^{2}Q^{2}+P^{2}\right)= \omega (a^\dagger a+\frac{1}{2 }) $ is the
 Hamiltonian of the system while   $\beta>0$ is the inverse temperature of the state $\rho_0$
 related to its average photon number
 $N=\mbox{Tr} \rho _{0}a^{\dagger }a$ by the identity
 $\beta  = \omega^{-1}\ln \frac{N+1}{N}$.
 For  $\mu \in \mathbb{C}$ (the set of all complex numbers) define the {\it displaced} version of $\rho_0$ as the state
\begin{eqnarray}
\rho _{\mu }=
D(\mu)\rho_{0}D^\dag(\mu)
\;, \label{shifted2}
\end{eqnarray}
which can be obtained, with good approximation, by pumping the mode
 with a laser far above threshold~\cite{gla63,YAMA}.
 In the above expressions  $\mu$ is the complex amplitude of the mode,  while
$D(\mu)\equiv \exp \left( a^{\dagger }\mu -a\bar{\mu}\right)$ is the displacement operator which
effectively describes the action of the laser.

One can consider the mapping $\mu \rightarrow \rho _{\mu }$ as a
c-q channel~(\ref{defCQ}) with continuous alphabet $\mathbb{C}$,
which encodes the signal $\mu $
into the quantum state $\rho _{\mu}$. From a
mathematical point of view it is the simplest channel describing
transmission of a classical signal on the background of additive
quantum Gaussian noise. This model can be naturally generalized to
describe many-mode (broadband) c-q channels \cite{cd,h02}
also in the presence of squeezing~\cite{hsh}.

\subsection{Multimode Bosonic systems}

A Bosonic system with $s$ degrees of freedom (modes) can be described in terms of
{\it canonical observables} $Q_{1},P_{1,}\cdots ,Q_{s},P_{s}$
 which satisfy the Heisenberg CCR
\begin{eqnarray}
[ Q_{j},P_{k}]=i \delta _{jk}\;,\qquad\quad  [Q_{j},Q_{k}]=[P_{j},P_{k}]=0\;.  \label{hei}
\end{eqnarray}
In quantum optics~\cite{gla63,YAMA} they coincide with
the system quadratures, and  are related to the Bosonic
creation-annihilation operators
via  the identities
$a_{j}^{\dagger }=\left( Q_{j}-i\omega_{j}P_{j}\right)/\sqrt{2 \omega_j}$,
$a_{j}=\left( Q_{j}+i\omega_{j}P_{j}\right)/\sqrt{2 \omega_j}$,
where $\omega _{j}$ are the frequencies of the modes.
Introducing the unitary Weyl operators $W(z)\equiv \exp[iR z]$,
where $R\equiv [Q_{1},P_{1,}\dots ,Q_{s},P_{s}]$ and $z$ is the column vector of real parameters $z=[x_{1},y_{1,}\dots ,x_{s},y_{s}]^{\top}$,
 the relations
(\ref{hei}) can be rewritten in the  equivalent Weyl-Segal form
\begin{eqnarray}
\fl \qquad W(z)W(z^{\prime })=
W(z+z^{\prime })\;\exp [\frac{i}{2}\Delta \left( z,z^{\prime }\right)]= W(z^{\prime })W(z)\; \exp[ i\Delta \left( z,z^{\prime }\right)]\;,
\label{WS}
\end{eqnarray}
where
\begin{eqnarray}
\Delta \left( z,z^{\prime }\right) \equiv \sum_{j=1}^{s}\left( x_{j}^{\prime
}y_{j}-x_{j}y_{j}^{\prime }\right) =z^{\top} \Delta  z^{\prime }\;,  \label{sym}
\end{eqnarray}
is the canonical {\it symplectic form} given by the skew-symmetric block matrix
\begin{eqnarray}
\Delta  \equiv  \mbox{diag}\left[
\begin{array}{cc}
0 & -1 \\
1 & 0
\end{array}
\right] _{j=1,\dots ,s}.
\end{eqnarray}
The $2s$-dimensional real vector space $\mathbb{R}^{2s}$ equipped with~(\ref{sym}) is
called the {\it symplectic space} and describes the classical phase space
underlying the quantum Bosonic system~\cite{ho1}.

The Weyl operators are closely related to
displacement operators by inducing the
translations  $W^\dag(-\Delta  z)  R  W(-\Delta  z)=R+z^\top$.
Moreover, any reasonable function of the canonical observables $\left\{
Q_{j},P_{j}\right\}$ is or can be approximated by a complex linear
combination of the Weyl operators $W(z)$. In particular,
 a state $\rho$ of the system  is uniquely described by its
{\it characteristic function} $\phi_{\rho }(z)$ defined by the relations
\begin{eqnarray} \label{defchar}
\phi_{\rho }(z)=\mathrm{Tr} \rho W(z)\;, \qquad \qquad
\rho = \int  \phi_{\rho }(z) W(-z)\frac{d^{2s}z}{(2\pi)^s} \;,
\end{eqnarray}
where in the second expression the integral is performed on the whole space $\mathbb{R}^{2s}$.
A state $\rho$ is called {\it Gaussian} if $\phi _{\rho }(z)$ has the
typical Gaussian form
\begin{eqnarray}
\phi _{\rho }(z)=\exp\left( im^{\top}  z-
\frac{1}{2}z^{\top}  \alpha  z\right) ,
\end{eqnarray}
where $m\in \mathbb{R}^{2s}$ and $\alpha $ is a real symmetric $2s\times 2s$
matrix. Similarly to the case of classical Gaussians, the components of $m^{\top}$
are mean values, i.e. the expectations  of the elements of  $R=[Q_{1},P_{1,}\dots
,Q_{s},P_{s}]$, and $\alpha$ is their (symmetrized) correlation matrix in the state $\rho$.
However in the quantum case $\alpha$ is not only nonnegative definite but
satisfies the matrix (Robertson's) uncertainty relation
\begin{eqnarray}
\alpha \geq \pm
i \Delta/2  \;.  \label{rob}
\end{eqnarray}
Examples of Gaussian states include vacuum, coherent, squeezed states and
also thermal states of oscillator systems, as well as their displacements
induced by action of a fixed source. For instance
the one-mode state (\ref{shifted2}) is Gaussian with the characteristic function
\begin{eqnarray}
\phi _{\rho _{\mu }}(z)=\exp \left[ \left( im_{q}x+m_{p}y\right) -\frac{
1}{2}\left( N+\frac{1}{2}\right) \left( x^{2}+y^{2}\right) \right] \;,
\end{eqnarray}
where $z=[x,y]^\top$,
%\tiny{\left[\begin{array}{c}x\\y\end{array}\right]}$,
$\mu = \frac{1}{\sqrt{2 \omega }}\left( \omega m_{q}+im_{p}\right)$, while Eq.~(\ref{rob}) amounts to $N\geq 0$, see e.g.~ \cite{ho1}.

\subsection{Quantum Gaussian channels}
\label{gauss_3} We now analyze those CPTP maps $\rho\rightarrow \rho
^{\prime }= \Phi \left[ \rho \right]$ which, when operating on
Gaussian input states,  preserve their structure. It turns out that
a convenient way to represent such mappings is to prescribe the
action of their  duals $\Phi^*$ (defined in Sec.~\ref{heisenbergp})
on the Weyl operators $W(z)$. One can show that the requirement that
the resulting dynamics induces a linear transformation of the
canonical observables of an open Bosonic system interacting with a Gaussian
environment imply
\begin{eqnarray}
\Phi^*[W(z)] = W(K z)\exp \left( il^{\top} z-\frac{1}{2}z^{\top}
\beta  z\right) \;,  \label{bosgaus}
\end{eqnarray}
where $l$ is the vector in $\mathbb{R}^{2s}$  while $K$ and $\beta$
are real $2s\times 2s-$matrices which satisfy the following
uncertainty relation
\begin{eqnarray}
\beta \geq \pm
i \left[ \Delta -K^{\top} \Delta K\right]/2 \;.  \label{nis1}
\end{eqnarray}
This inequality is a {\it necessary and sufficient} condition for
the complete positivity of $\Phi$~\cite{dvv,hw,ew}. Thus, a Gaussian
channel is completely characterized by the three quantities $ \left(
K,l,\beta \right) ,$ where $\beta $ satisfies (\ref{nis1}) (note
that there is no restriction on $l$). As any completely positive
map, a Gaussian channel admits Kraus decomposition, explicitly
described in \cite{simon} in the case of one mode.

By using Eqs.~(\ref{Vheisenberg}) and (\ref{defchar}) one can verify
that the mean $m$ and the correlation matrix $\alpha$ of an input
state $\rho$ are  transformed by the channel according to the
relations
\begin{eqnarray}
m\rightarrow m^{\prime }=K^{\top} m+l\;,\qquad \quad \alpha
\rightarrow \alpha ^{\prime}=K^{\top} \alpha  K+\beta \;,
\end{eqnarray}
which, due to Eq.~(\ref{nis1}), guarantee that the output counterparts of the Gaussian inputs~(\ref{defchar})
are again Gaussian states.

A special subclass of the Gaussian channels defined above is
obtained for $\beta=0$. Under this constraint Eq.~(\ref{nis1})
forces the matrix $K$ to be {\it symplectic}, i.e. to fulfill the
condition
\begin{eqnarray} \label{sympl}
K^{\top} \Delta  K=\Delta\; .
\end{eqnarray}
Equation~(\ref{bosgaus}) implies that these special channels induce
the following linear transformation of the canonical observables
\begin{eqnarray}
R\rightarrow R^{\prime } = \Phi^*[ R]  =R K +l^{\top}\;, \label{canonical}
\end{eqnarray}
whose components still satisfy the CCR (\ref{hei}) due to
Eq.~(\ref{sympl}) (for this reason the mapping~(\ref{canonical}) is
called {\it canonical}). These  transformations represent a unitary
evolution of the system which, in the Schr\"{o}dinger
representation, can be expressed as a concatenation of two
elementary processes
in which one first shifts the coordinates of the system, and then
applies a proper symplectic transformation. The first process
amounts to the unitary transformation induced by the Weyl operator
$W(-\Delta l)$, i.e. $\Phi[\rho]= W(-\Delta  l) \rho W^\dag(-\Delta
l)$. The second process can be described by a unitary transformation
$U_K$ in ${\cal H}$ (quantization of the symplectic matrix $K$) such
that~\cite{ho1}
\begin{eqnarray}
U_{K}^{\dag }W(z)U_{K}=W(K z)\;,\qquad \mathrm{hence\qquad }U_{K}^{\dag
}RU_{K} =R K\;.
\end{eqnarray}
Any reversible dynamics of Bosonic system which in the Heisenberg
picture is linear in the canonical observables as in
Eq.~(\ref{canonical}) can be described in these terms. This is the
case for the unitary transformations $\exp[iH]$ induced by
Hamiltonians $H$ which are at most quadratic polynomials of the canonical
observables, such as the free Hamiltonian of a set of harmonic
oscillators, or the Hamiltonian governing a linear amplifier optical
process~\cite{YAMA}.

The presence of irreversible, noisy dynamics for Gaussian channels
is signaled by having $\beta\neq 0$ in Eq.~(\ref{bosgaus}). In this
case it is possible to produce a unitary
representation~(\ref{ivolu}) of $\Phi$ in terms of a unitary
coupling with a multimode  Bosonic environment $E$ characterized by
the canonical observables $R_E$ that induces the following linear
input-output relation
\begin{eqnarray}
\fl \qquad \qquad R \rightarrow R^{\prime }=R K+R_{E} K_{E}\;, \qquad  \quad
R_{E} \rightarrow R_{E}^{\prime }=R L+R_{E} L_{E}\;,\label{in-out}
\end{eqnarray}
where $T=\tiny{\left[
\begin{array}{cc}
K & L \\
K_{E} & L_{E}
\end{array}
\right]}$ is a symplectic $2\left( s+s_{E}\right) \times 2\left(
s+s_{E}\right)$ matrix ($s_E$ is the number of environmental modes
entering the unitary representation). In this framework the initial
state $\rho_E$ of the environment $E$ is chosen to be
Gaussian~\cite{cegh,cegh1} with the correlation matrix $\alpha _{E}$
satisfying $\beta =K_{E}^{\top} \alpha _{E} K_{E}$ (the minimum
value of $s_E$ under the assumption that $E$ is initialized in a
pure state computed in Ref.~\cite{cegh1} is found to be equal to the rank of
the matrices $\left[ \beta \pm i (\Delta -K^{\top} \Delta
K)/2\right]$).
The second relation in Eq.~(\ref{in-out}), depending on the
(non-)purity of $\rho_E$, describes the (weakly) complementary
channel $\rho \rightarrow \rho_{E}^{\prime }$ of $\Phi$ which is also a Gaussian channel.

\subsection{General properties of the Gaussian channels} \label{gpgc}
Before discussing some specific example of Gaussian channels in the
next Section we shall list their general properties \cite{hw}:

\begin{enumerate}
\item Gaussian states are transformed into Gaussian states.

\item The dual of a Gaussian channel transforms a polynomial in the canonical
variables into polynomial of the same degree.

\item The concatenation of Gaussian channels is again a Gaussian channel. In fact, let $
\Phi _{j};j=1,2,$ be two Gaussian channels characterized by the
parameters $(K_{j},l_{j},\beta _{j})$, then by using
(\ref{bosgaus}), the composite map $\Phi _{2}\circ \Phi _{1}$ is
Gaussian  with parameters
\begin{eqnarray}
K &=&K_{1} K_{2}\;,  \nonumber  \\
l &=&K_{2}^{\top} l_{1}+l_{2}\;,  \label{superp} \\
\beta &=&K_{2}^{\top } \beta _{1} K_{2}+\beta _{2}\;.  \nonumber
\end{eqnarray}

\item Gaussian channels are covariant under the action of Weyl operators. That is
for all $\rho$ one has
\begin{eqnarray}
\Phi [ W^\dag(z) \; \rho \;W(z)]=W^\dag (K^{\prime } z)\; \Phi [
\rho]\;W(K^{\prime } z)\;,  \label{gauscov}
\end{eqnarray}
where $K^{\prime }=\Delta ^{-1} K^{\top}\Delta$. Due to this
property, the displacement parameter $l$ entering~(\ref{bosgaus}) can always be
removed as it can be compensated by a proper unitary transformation
at the input or at the output of the communication line.

\item If $K$ is invertible, then $\Phi [ I]=|\det K|^{-1}I$, in
particular, $\Phi $ is unital if and only if $|\det K|=1$~\cite{heg}. If $\Phi$
is Gaussian channel with parameters $(K, l, \beta)$, then $|\det
K|\Phi^*$ is Gaussian channel with parameters $(K^{-1}, -(K^{-1})^\top l,
(K^{-1})^\top\beta K^{-1})$ (for channels in one mode this duality was observed in ~\cite{simon}).

\item Let $\Phi $ be quantum Gaussian channel with parameters $(K,l,\beta). $
It is entanglement-breaking if and only if $\beta$ admits the
decomposition \cite{hebr}
\begin{eqnarray}
\beta =\alpha +\nu\; ,\quad \mathrm{where}\quad \alpha \geq
\frac{i}{2}\Delta\; ,\quad \nu \geq \frac{i}{2}K^{\top} \Delta  K\;.
\label{decom1}
\end{eqnarray}
In this case $\Phi $ admits a  representation of the form
\begin{eqnarray}
\Phi [ \rho ]=\int\; W(z) \sigma _{B}W^\dag(z) \; p_{\rho
}(z)d^{2s}z\;,
\end{eqnarray}
where $p_\rho (z)$ is the probability density in the state
$\rho$ of a Gaussian measurement with outcomes $z$, followed by the
displacement $W(z)$ of a Gaussian state $\sigma _{B}$ with
covariance matrix $\alpha$. Gaussian measurements are those POVMs
which transform quantum Gaussian states into Gaussian probability
distributions on $\mathbb{R}^{2s}$, see e.g.~\cite{ho1} for detail.
In quantum optics these are implemented by optical homo- and/or
hetero-dynes combined with linear multiport
interferometers~\cite{YAMA}.
\end{enumerate}

\subsection{The case of one mode} \label{onemode}

For ``continuous variables'' systems the one-mode channels ($s=1$)
play a role similar to qubit channels for finite systems. Therefore
it is interesting and important to have their classification under
unitary equivalence with respect to canonical
transformations~(\ref{canonical}),  the problem a solution for which is given in~\cite{hclass,cgh}.
In what follows we consider the most relevant
examples of those maps whose efficiency in transferring information will be
presented in Sec.~\ref{gauscapacity}.
\begin{enumerate}
\item {\it Attenuation} channels. These are characterized by
transformation~(\ref{bosgaus}) with $l=0$ and
\begin{eqnarray}
K=k\small{\left[
\begin{array}{cc}
1 & 0 \\
0 & 1
\end{array}
\right] }\;,\qquad \quad \beta = \left(
N_{0}+\frac{|1-k^{2}|}{2}\right) \small{\left[
\begin{array}{cc}
1 & 0 \\
0 & 1
\end{array}
\right]} \;, \label{param}
\end{eqnarray}
where $k\in (0,1)$ and $N_0$ is nonnegative -- the latter being the
condition which enforces the inequality~(\ref{nis1}).  The resulting
mapping can be described in terms of a coupling~(\ref{ivolu}),
mediated by a beam-splitter of transmissivity ${k}$~\cite{cd,YAMA},
between the input state of the system and an external Bosonic
environmental mode initialized in a thermal (Gibbs) state
(\ref{qupo})  with the mean photon number $N=N_0/|1-k^{2}|$.

\item
{\it Ampification} channels. These are described by the same
matrices $K$ and $\beta$ of Eq.~(\ref{param}) were now however the
parameter $k$ assumes values larger than~$1$. As in the previous
case the resulting mapping can be described in terms of a linear
coupling with an environmental state initialized in a thermal state
with photon numbers $N_0$. In this case however the coupling is
provided by a two-mode squeezing Hamiltonian which induces a linear
amplification of the impinging field~\cite{YAMA}.

\item {\it Additive classical noise} channels. These maps are characterized by
the parameters
\begin{eqnarray}
K=\small{\left[
\begin{array}{cc}
1 & 0 \\
0 & 1
\end{array}
\right]} \;,\qquad \quad \beta = N_{0}\small{\left[
\begin{array}{cc}
1 & 0 \\
0 & 1
\end{array}
\right] }\;,
\end{eqnarray}
with $N_0\geqslant 0$. In this case it is possible to write a simple
integral representation for the channel
\begin{equation}\label{anc}
\Phi \left[ \rho \right] =\frac{1}{\pi N}\int D(\zeta )\rho D^\dag(\zeta)
\; \exp \left( -{|\zeta |^{2}}/{N}\right) d^{2}\zeta \;,
\end{equation}
with $D(\zeta)=\exp[\zeta a^\dag - \zeta a]$,
which is a continuous analog of the Kraus decomposition expressing the fact that
the channel performs random Gaussian displacements of the input state as a result of action of a classical random source,
e.g. see Ref.~\cite{HALL}. These maps has been also analyzed in the context of universal cloning machines~\cite{LINDCLO,cerfclo}.
\end{enumerate}

Let us now see which of the above channels are entanglement-breaking basing
on the decomposability criterion (\ref{decom1}).
To do so we rely upon the simple fact that
$(N+{1}/{2})I\geq i\Delta/2$ if and only if $N\geq 0.$ Therefore since we have $K^{\top} \Delta
K=k^{2} \Delta$, it follows that in this case the decomposability condition holds if and only if
$\beta \geq \frac{i}{2}(1+k^{2})\Delta$ which is equivalent to
$N_0+{|1-k^{2}|}/{2}\geq {(1+k^{2})}/{2}$ or
\begin{eqnarray}
N_0\geq \min \left( 1,k^{2}\right) .  \label{doma}
\end{eqnarray}
This gives the condition for entanglement breaking applicable to the
channels of the classes (i), (ii) and (iii).

\section{Entropy, information, channel capacities}
\label{entropic}

The most profound results making the essence of information theory
-- coding theorems --  have asymptotic nature and concern
 the transmission of  messages formed by arbitrarily long, ordered
sequences of symbols which are emitted by a  ``source''  (say a radio station).
In practical applications the index that enumerates the various entries of the message represents  either
the (discrete) time when they appear  sequentially in the
communication process, or the number of ``modes'' in which a given
signal can be processed simultaneously in parallel (e.g. the
frequency modes in which a radio wave can be decomposed, or the
cells that form a memory register). In a statistical description  a
question inevitably arises regarding what model for correlations
between the different symbols of a message should be accepted. The
simplest yet basic is the {\it memoryless} model where both the
production of  the symbols emitted  by the source, and their
subsequent transformations associated with the propagation through a
communication line, are statistically identical and independent. The
coding theorems of information theory are not at all trivial already
for this  case. At the same time they give a basis for considering
more complicated and realistic scenarios taking into account memory
effects. Moreover, the effects of entanglement are demonstrated in the memoryless case in the most spectacular way.
Therefore in our presentation we concentrate on the
memoryless configurations, providing, where appropriate, references
for the more advanced memory models.

In the following we start reviewing some basics results of the classical information theory that allow us to
evaluate the information content of a classical message (Shannon's first coding theorem) and to introduce the
notion of  {\it capacity} for a  communication channel (Shannon's second coding theorem).
Moving into the quantum domain we will face then the fundamental issue on how
to generalize these results  in order to have a proper quantum mechanical measure
of information. The task is particularly challenging due to the complex nature of
quantum information (see discussion in Sec.~\ref{sec:ClassicalQuantum}).
In particular, there is a fundamental conceptual distinction between the amount of {\it classical
information} that it is possible to store in a quantum system, and the amount of {\it quantum
information} that it can accommodate. At the level of communication theory, this forces to
introduce different alternative generalizations of the Shannon capacity, see e.g.~\cite{bs,shorreview}.

\subsection{Information transmission over a classical channel} \label{transclas}

Consider a classical  source of information which emits $n$-long
sequences $w=(x_{1},\dots ,x_{n})\in {\cal X}^n$  formed by symbols
$x_j$ that are randomly extracted from an alphabet ${\cal X}$
according to a distribution $P=\{p_x\}$ (the process being repeated
{\it identically} and {\it independently} for each element of the
sequence as mentioned previously). In classical information theory
the amount of information that is contained in one of those
sequences is measured by the minimal number of binary digits (bits)
necessary for its  binary representation (coding).  The first
Shannon coding theorem~\cite{shannon}  says  that this number is
$\approx{nH(X)}$, where $H(X)$ is the entropy~(\ref{defshannon}) of
the distribution~$P$ that characterizes the source. This is obtained
by showing that for $n\gg 1$  the {\it typical} messages $w$ which
are most likely to be produced by the source all have approximately
equal probabilities $2^{-nH(X)}$, so that their number is $\approx
2^{nH(X)}$. Since the number of bits necessary to enumerate such
{\it typical} messages is $\approx n H(X)$, it follows that, by
tolerating an asymptotically small error probability, one can use
this same number of bits to encode {\em all} possible messages
emitted by source (e.g. associating to all the non typical ones  a
fixed erasure symbol). This result can be extended
also to a wide class of stationary correlated sources
models~\cite{cover} and provides an operational interpretation of
$H(X)$ as the information content per symbol of the  average message
emitted by the source.

Assume now that the messages produced by the source are transmitted
through a classical noisy channel which maps the $n$-long input
sequences $w=(x_{1},\dots ,x_{n})$ into $n$-long output messages
$w'=(y_1,\dots,y_n)\in {\cal Y}^n$ whose symbols $y_j\in {\cal Y}$
appear with  output probability distribution  $P^{\prime
}=\{p_{y}^{\prime}\}$ where $p_{y}^{\prime}=\sum_{x}p(y|x)p_{x}$.
 In this expression $p(y|x)$ is the single letter
conditional probability distribution which fully characterizes the
process in the memoryless scenario. Memory channels  in which the noise tampering with
the transmission process acts on the letters of the message in a
correlated fashion, require instead to assign the complete joint
input-output probabilities for the messages, see e.g.~\cite{GALL}.

Due to the stochastic nature of the transformation $w\rightarrow
w'$, the distinguishability of the various input messages is not
necessarily preserved. Still one can show that by  properly
selecting the codebook of messages $w$ (encoding), it is possible to convey (reliably) a
definite amount of information to the output of the transmission
line. Loosely speaking, the second Shannon coding theorem establishes that the
amount of information per transmitted symbol that can be recovered
from the channel outcomes is given by the {\it Shannon capacity}
\begin{eqnarray}
C_{Shan} \equiv \max_{X}I(X;Y)\;,  \label{CC}
\end{eqnarray}
where the maximum is taken over all possible distributions $P$ of
the input $X$. In this expression $I(X;Y)$ is the {\it mutual} (or
{\it Shannon}) {\it information},
\begin{eqnarray}
I(X;Y)\equiv H(Y)-H(Y|X)=H(X)+H(Y)-H(X,Y)\;,  \label{DefSI}
\end{eqnarray}
with $H(Y)$ being the entropy of  the output distribution $P^{\prime
}=\{p_{y}^{\prime}\}$ which measures the total information content
of the output of the channel, both useful -- due to the uncertainty
of the signal $X$, and harmful -- due to the noise in the channel.
Similarly, $H(X,Y)=-\sum_{x,y}p_{x,y}\log_2 p_{x,y}$ is the
{\it joint entropy} of the pair of random variables $X,Y$ whose
joint distribution is $p_{x,y}=p_x p(y|x)$. Finally $H(Y|X)$ is the
{\it conditional entropy} (called also {\it information
loss}) defined as
\begin{eqnarray} \label{conditional}
H(Y|X) &\equiv &\sum_{x}p_{x}H(Y|X=x) =-\sum_{x}p_{x}\sum_{y}p(y|x)\log_2 p(y|x) \\
&=&H(X,Y)-H(X)\;,
\end{eqnarray}
which reflects the effect of the {\it noise} in the communication.

More precisely, the second Shannon theorem says that by using special (block) encoding
 of the messages at the input and decoding at the output of the
channel it is possible to transmit $\approx 2^{nC_{Shan}}$
($n\rightarrow \infty )$ messages with asymptotically vanishing
error, and it is impossible to safely transmit more whatever
encoding and decoding are used. Thus $C_{Shan}$ is the ultimate
asymptotically errorless transmission rate for the channel $p(y|x)$.
In practice the optimal encoding and decoding doing the miraculous
job of making the noisy channel partially invertible, i.e.
transparent to some selected messages, are extremely difficult to
approach, but $C_{Shan}$ gives the benchmark describing the
exponential size of such an ideal codebook. Another insightful
interpretation of the quantity $C_{Shan}$ is given by the ``reverse
Shannon theorem'' (formulated and proved surprisingly rather
recently~\cite{bsst2}, under the influence of ideas from quantum
information theory). This theorem implies that  the ratio $\log_2 |\mathcal{X}|/C_{Shan}$ is
equal to the number of copies of the noisy channel $p(y|x)$, needed
to simulate (asymptotically) the ideal (noiseless) channel with the
maximal capacity $\log_2 |\mathcal{X}|$.

The memoryless nature of the
classical channel we are considering here is reflected by the additivity property
\begin{eqnarray}\label{ADDc}
C_{Shan}^{(n)} = n\; C_{Shan}\;,\end{eqnarray} where $C_{Shan}^{(n)}$
 is the Shannon capacity of the
$n$-block channel one obtains by treating
$n$ instances of the source as a new source that emits $n$-long strings $w=(x_1,\cdots, x_n)$ as individual super-symbols. Accordingly in the definition of $C_{Shan}^{(n)}$ the maximum  in (\ref{CC}) is
taken over the all input distributions $P^{(n)}$ on the space ${\cal X}^n$, including the correlated ones.

\subsection{Quantifying information in a quantum world}

Let us start with a {\it quantum
source} of information that produces $n$-long sequences of quantum symbols (pure
states) $|\psi_j\rangle\in {\cal H}$, creating  factorized states of the form
$|\Psi\rangle = |\psi_{j_1}\rangle \otimes \cdots \otimes |\psi_{j_n}\rangle \in {\cal H}^{\otimes n}$
(a practical example is a pulsed laser that emits series of random optical signals characterized
by different intensity values).
As in the classical case we assume that each symbol
composing the sequence $|\Psi\rangle$ is produced in a memoryless way, i.e.
it is extracted from a given set of possible states independently with certain probability distribution $\{p_j\}$.
We ask what is the minimum number of qubits per symbol necessary to express the generic state
produced by such a process.
Schumacher and Josza~\cite{js} answered this question
by showing  that in the limit  ($n\to\infty$) this number
coincides with the von Neumann entropy $S(\rho)$
of the density operator $
\rho =\sum_{j}p_{j}|\psi _{j}\rangle \langle \psi _{j}|$ which represents the average state emitted by the source.
More precisely,  they proved that with the high (asymptotically unit) probability,  the sequences $|\Psi\rangle$
span a subspace of ${\cal H}^{\otimes n}$
of dimensionality $\approx 2^{nS(\rho)}$.
Since the density matrix describing the average sequence $|\Psi\rangle$   is
$\rho^{\otimes n}$, this can be rephrased by saying that ${nS(\rho)}$ is the logarithmic
size of a ``quantum register'' in which a given {\it quantum
message} $\rho^{\otimes n}$ can be ``packed'' optimally with negligible loss of
information. Thus in quantum information theory the logarithm of the dimensionality
of the space of state vectors carrying information is the measure of
information content of the system, and plays a role similar to the
logarithm of size of the codebook for classical messages.
Analogously to what the first Shannon coding theorem says for the Shannon entropy,
the above result provides an operational characterization for the von Neumann entropy
$S(\rho)$, presenting it as a fundamental measure of
 the amount of quantum information that can be stored in the density matrix~$\rho$.

This result admits also another operational interpretation. Viewing the quantum source defined above as a
special instance of a c-q channel~(\ref{defCQ}) that encodes the classical variable $j$ into the pure quantum states $|\psi_j\rangle$, the result of Ref.~\cite{js}  indirectly
quantifies the amount of quantum resources (qubits)
that are necessary to carry on such an encoding. The generalization of this result to arbitrary c-q channels is the subject of the next Section.

\subsection{The classical capacity of quantum channel: part I}\label{ccqc1}

In Sec.~\ref{q-c-q}  we have seen that
the simplest quantum model of a communication line is the c-q channel~(\ref{defCQ})
where some classical data $x\in{\cal X}$ are mapped
into a fixed family of the output quantum states $\{\rho _{x}\}$ in the
receiver space $\mathcal{H}$.
If the letters of the message $w=(x_{1},\dots ,x_{n})$ are transmitted independently of
each other (no memory) then at the output of the composite channel
one has separable state $\rho _{x_{1}}\otimes \dots \otimes \rho_{x_{n}}$
in the space~$\mathcal{H}^{\otimes n}$. Decoding at the output requires a
quantum measurement in $\mathcal{H}^{\otimes n}$, the outcome of
which gives an estimate $w'$ for $w$.
For each specific choice of such measurement we can thus define a classical channel
which takes the classical random variable $w$ to its output counterpart $w'$ (passing through
the quantum stage  $w \rightarrow \rho _{x_{1}}\otimes \dots \otimes \rho_{x_{n}}$).

Its ability in transferring  classical information can now be evaluated by the
associated Shannon capacity defined as in Sec.~\ref{transclas}.
Define $C^{(n)}$ to be the maximum of this quantity obtained by optimizing it with respect to
all possible measurements on $\mathcal{H}^{\otimes n}$.
It turns out that in general,
due to existence of entangled measurements at the output,
differently from~(\ref{ADDc}), one may have
$C^{(n)}>n\; C^{(1)}$. In other words, for the c-q memoryless
channels the transmitted classical information can be {\it strictly
superadditive}. A proper definition of the capacity requires hence
a {\it regularization} with respect to the block coding size, i.e.
\begin{eqnarray}
C_{\chi}=\lim_{n\rightarrow \infty }C^{(n)}/n\;.  \label{cc}
\end{eqnarray}

Remarkably, the quantity $C_{\chi}$, defined rather implicitly by this equation, admits an explicit entropic expression.
For a  statistical ensemble consisting of the density matrices
$\{\rho_x\}$ with the probabilities $\{p_x\}$ define the $\chi$-{\it information} as
\begin{eqnarray}
\chi \equiv \chi \left( \{p _{x}\}, \{\rho _{x}\}\right) =S\left(
\sum_{x}p_{x}\rho _{x}\right) -\sum_{x}p_{x}S(\rho _{x})\;. \label{chiq}
\end{eqnarray}
This quantity is nonnegative due to the concavity of the von Neumann entropy (see also Sec.~\ref{relentropy}).
To certain extent the $\chi$-information can be regarded as a quantum analog of the Shannon information  defined in Eq.~(\ref{DefSI}) (see however the discussion in Sec.~\ref{entexch}). Both these quantities are the differences between the overall output entropy of the channel and a term which can be interpreted as the conditional entropy (loss). Furthermore,  $\chi$
provides a fundamental upper bound, first proved in~\cite{hol73}, for the Shannon information  $I(X,Y)$  between the random variable $X$
having the distribution $\{p _{x}\}$ and the random variable $Y$ describing the outcome
of a measurement at the output of the channel aimed to recovered the value $x$, namely
\begin{eqnarray}
I(X,Y)\leq \chi \left( \{p _{x}\}, \{\rho _{x}\}\right) \;.
\label{ebound}
\end{eqnarray}

The coding theorem established by Holevo~\cite{h96,hol79} and by
Schumacher-Westmoreland~\cite{sw97} shows that there
exist block coding strategies which, in the limit of large $n$, saturate
the bound imposed by Eq.~(\ref{ebound}). Thus the abstractly defined
capacity~(\ref{cc}) acquires the following compact ``one-letter'' expression
\begin{eqnarray}
C_{\chi}=\max\limits_{\{p_{x}\}}\;\;
 \chi \left( \{p _{x}\},\{ \rho _{x}\}\right)
\;, \label{ct}
\end{eqnarray}
(the maximization is
 performed over probabilities $\{p_x\}$ while keeping fixed $\{\rho_x\}$).
 This relation  can be regarded as the
``classical-quantum'' analog of the second Shannon coding theorem for
the noisy channel.  For the later modifications of its proof see~\cite{winter99,OGAWA,OGNA,HayaNaga,sgm}.

An apparent but important consequence of the bound (\ref{ebound}) is the inequality
\begin{eqnarray}
C_\chi \leq \log_2 \dim \mathcal{H}\;,  \label{ct1}
\end{eqnarray}
in which the equality is attained if the quantum source operates with orthogonal states $\rho_x$.
Thus, the fact that the space $\mathcal{H}$ contains
infinitely many state vectors, does not allow to increase the classical
capacity above the ultimate information resource of the quantum
system; increasing the number of signal vectors forces them to
become nonorthogonal and hence less and less distinguishable. This is in line with the observation
in the previous section where it was shown that
the logarithm of dimensionality of the Hilbert space of a quantum system  determines
the ultimate bound on the amount of quantum information one can store in it.
\newline

\textbf{Example 1:} Consider the binary input signal $x=\pm 1$, and let $\rho
_{\pm 1}$ be the coherent state of a monochromatic laser beam with the
complex amplitude $\pm z$. This defines  a c-q channel with two
pure nonorthogonal states~\cite{h} whose classical capacity~(\ref{ct}) can be computed as
\begin{eqnarray}
C_\chi=h_{2}\left( \frac{1+\epsilon }{2}\right) ,  \label{C}
\end{eqnarray}
where $\epsilon =\langle z|-z\rangle =\exp (-2|z|^{2})$ is the overlap between the two coherent states,
while
$h_2(p)=-p\log_2 p-(1-p)\log_2 (1-p)$ is the Shannon binary entropy. Also  the quantity
$C^{(1)}$ can be explicitly computed \cite{levi} yielding
\begin{eqnarray}
C^{(1)}=1-h_{2}\left( \frac{1+\sqrt{1-\epsilon ^{2}}}{2}\right) \;.  \label{C1}
\end{eqnarray}
For this special case one can then easily verify that $C_\chi/C^{(1)}>1$, the ratio
 tending to $\infty$ in the limit of the weak signal, as $\epsilon \rightarrow 1$, i.e. $
z\rightarrow 0$.
\newline

\textbf{Example 2: } Consider the c-q channel
in which the classical alphabet  ${\cal X}$ is composed by three symbols (say $x=0,+,-$),
which are mapped into  three
equiangular pure states  $\ket{\psi_x}$ of a qubit system,
\begin{equation} \label{trine} \ket{\psi_0}=\left[
\begin{array}{c}
  1 \\
  0 \\
\end{array}
\right]\;,\quad\ket{\psi_\pm}=\left[
\begin{array}{c}
  -1/2 \\
  \pm \sqrt{3}/2 \\
\end{array}
\right]\;.
\end{equation}
 In this case of
equal probabilities, $p_x=1/3$, the average density matrix coincides with the
chaotic state (i.e.
  $\rho=  \sum_x |\psi_{x}\rangle\langle \psi_x|/3 =I/2$).
The classical capacity~(\ref{ct})  of the channel is hence $C_\chi=S\left(I/2\right)=1$
which saturates the bound (\ref{ct1}) in spite of the fact that
the states are not orthogonal.  Also for this channel the gap between $C_\chi$ and $C^{(1)}$ can be shown. One has
$C^{(1)}\approx 0.645$, the value attained for
the distribution $ p_{+}=p_{-}=\frac{1}{2},\ p_{0}=0$ and for the
measurement corresponding to the orthonormal basis, optimal for
discrimination between the two equiprobable states
$\ket{\psi_x},x=\pm$ \cite{sas,shad}. However the optimal encoding and
decoding for this case are unknown, as for the most of the Shannon
theory.

More generally, let $\{\ket{\hat{\psi}_x}; x\in\mathcal{X}\}$
be an overcomplete system (\ref{overcomp}) in a $d$-dimensional space (the above example being
just a special case with  $\ket{\hat{\psi}_x} =\sqrt{2/3}\ket{\psi_x}$).
Then the overcompleteness relation  can be written
as $\sum_x p_x \rho_x=I/d$,
where $p_x=\braket{\hat{\psi}_x}{\hat{\psi}_x}/d$ and
 $\rho_x={\ketbra{\hat{\psi}_x}{\hat{\psi}_x}}/{\braket{\hat{\psi}_x}{\hat{\psi}_x}}$.
This implies that the c-q channel $x\rightarrow\rho_x$ has the capacity $C_\chi=S(I/d)=\log_2 d$.
Furthermore, since the inequality in (\ref{ebound}) is strict unless the operators $\rho_x$ commute~\cite{hol73},
one can conclude that $C^{(1)}<C_\chi = \log_2d$ unless $\{\ket{\hat{\psi}_x}\}$ is an orthonormal basis.
\newline 

{\bf Example 3:} Consider the  c-q channel model introduced in Sec.~\ref{onemodeexa} where
 a continuous  alphabet $\mathbb{C}$ is encoded into quantum states via the mapping  $\mu \rightarrow \rho _{\mu}$  with
$\rho _{\mu}$ given by (\ref{shifted2}).
If one tries to compute the classical capacity of
such a channel by a continuous analog of the formula~(\ref{ct}) where the probabilities
$p_x$  are replaced by probability distributions $p(\mu)$ on the complex plane  $\mathbb{C}$, i.e.
\begin{eqnarray}
C_\chi =\max_{p(\mu )}\left\{ S\left( \int p(\mu )\rho _{\mu}\; d^{2}\mu \right) -\int
p(\mu )\; S(\rho _{\mu})\; d^{2}\mu \right\}  \;,\label{sup}
\end{eqnarray}
one gets infinite value as it is to be expected for a channel with infinite
input alphabet. To obtain a reasonable finite result one should introduce a
constraint onto possible input distributions $p(\mu)$ restricting the ``energy'' that one can actually
add in the fiber, namely
\begin{eqnarray}
%\frac{1}{2 }
\int |\mu|^{2}\,p(\mu )d^{2}\mu \leq E\;.  \label{3-5}
\end{eqnarray}
With this constraint, the maximum (\ref{sup}) can be evaluated by taking into
account two facts: first, the states (\ref{shifted2}) are all unitarily
equivalent and have the same entropy
\begin{eqnarray}\label{defig}
S(\rho _{\mu})=S(\rho _{0})=g(N)\equiv (N+1)\log_2 (N+1)-N\log_2 N  \;,
\end{eqnarray}
see e.g. \cite{ho1}; second, the constraint (\ref{3-5}) implies
\begin{eqnarray}
\mbox{Tr}\,\bar{\rho} \;a^{\dagger }a\leq N+E\;,  \label{3-6}
\end{eqnarray}
where $\bar{\rho}=\int p(\mu )\rho _{\mu}d^{2}\mu$ is the average quantum state transferred in the communication. Consequently, by the
maximal entropy principle, the entropy $S\left( \bar{\rho}\right)$ is
maximized for the Gaussian distribution $p(\mu )=\frac{1}{\pi E}\exp\left( -{|\mu |^{2}}/{E}\right)$ giving the value $g(N+E)$, whence the channel
capacity is
\begin{eqnarray}
C_\chi =g(N+E)-g(N)\;.
\end{eqnarray}

In the classical limit of large energies
(i.e. $N\rightarrow \infty$, $E/N\rightarrow \mathrm{const}$)
this turns into $C_\chi =\log_2 ( 1+{E}/{N})$, which can be regarded
as a generalization of the famous Shannon's formula~\cite{shannon}
$C_{Shan}=\frac{1}{2}\log_2 \left( 1+{E}/{N}\right)$,
for the capacity of memoryless channel with additive Gaussian white noise of
power $N$ (the factor $1/2$ being absent in the quantum case because one degree
of freedom amounts to the two independent identically distributed real
amplitudes). The capacity formula can be generalized to many-mode (broadband) c-q
Gaussian channels \cite{cd,h02}
also in the presence of squeezed seed states~\cite{hsh}.

In the Shannon theory one considers memory channels with stationary Gaussian noise
by making spectral decomposition of the time series in question. It turns out that
the modes corresponding to different frequencies can be considered asymptotically independent if
the observation time is very large. Then effectively one has a set of parallel independent channels which
can be approached as a kind of ``memoryless'' composite channel \cite{cover}. Similar reduction to the
memoryless channel in the frequency domain can be elaborated for the c-q channels with additive stationary quantum Gaussian noise~\cite{h02}.

\subsection{The classical capacity of quantum channel: part II} \label{partII}
In the previous section we focused on
the basic case of quantum channels that can be
used to transfer classical messages, the c-q channels. They are formally constructed by
fixing a set of possible quantum letters $\rho_x$ in the space ${\cal H}$ organized in
the $n$-long sequences  $\rho _{x_{1}}\otimes \dots \otimes \rho_{x_{n}}$ that form
the codewords in which the classical information can be encoded.

More generally, assume that the sender of the classical information is allowed
to use as input {\it any} (possibly entangled~\cite{xbennet,bs}) joint density matrix
$\rho^{(w)} \in \mathfrak{S} ({\cal H}^{\otimes n})$ of $n$ quantum information carriers.
Here $w$ denotes a classical message encoded into the state $\rho^{(w)}$. We assume that
there are $N$ different messages to be transmitted, hence $w$ can be simply the number
of the message, $w=1,\dots,N$.
Assume also that during the transmission stage each component of the
codeword $\rho^{(w)}$  is individually affected  by the same noisy channel (memoryless regime),
producing the outputs of the form
\begin{eqnarray}
\rho^{(w)} \qquad \longrightarrow \qquad \Phi^{\otimes n}[\rho^{(w)}]\;,\label{cqmapping}
\end{eqnarray}
where $\Phi^{\otimes n}$ is the $n$-fold  tensor product defined similarly to~(\ref{compo1}) of
the quantum channel $\Phi$ which gives the statistical description~(\ref{ivolu})  of the
interaction of a single carrier with the environment (i.e. noise).
Notice that for each given collection  of input  density matrices $\{ \rho^{(w)}; w=1,\dots,N\}$, Eq.~(\ref{cqmapping})  defines an effective  c-q channel
whose quantum letters are the density matrices $\Phi^{\otimes n}[\rho^{(w)}]$ .
The associated classical capacity  of such c-q channel is then computed along the lines of Sec.~\ref{ccqc1},
\begin{eqnarray}
C_{\chi}(\Phi^{\otimes n})= \max_{\{p_{w}\},\{\rho^{(w)}\}}
\chi\left(\left\{p_w \right\},\left\{\Phi^{\otimes n}[ \rho^{(w)}] \right\}\right)\;,
\label{cchi}
\end{eqnarray}
where now, given the freedom the sender has in selecting the input density matrices of the channel,
the maximum is performed over the set of statistical ensembles formed by the
probabilities $\{p_{w}\}$ {\it and} the states $\{\rho^{(w)}\}$.
Loosely speaking, Eq.~(\ref{cchi})  gives the ultimate rate of classical bits that one can transmit when using quantum block-letters of size $n$, hence corresponding
to the rate  $C_{\chi}(\Phi^{\otimes n})/n$ bits per individual  use of the channel $\Phi$.
Therefore the ultimate (asymptotically achievable) rate is given by the expression
\begin{eqnarray}
C(\Phi )=\lim_{n \rightarrow \infty }\frac{1}{n}C_{\chi }(\Phi^{\otimes n})\;.
\label{coding}
\end{eqnarray}
It should be stressed  that the regularizations involved in Eqs.~(\ref{coding}) and (\ref{cc}) have different origin. While
taking the limit in Eq.~(\ref{cc}) is required by superadditivity due to possible use of entanglement at the decoding
stage of the communication process (see previous Section),
the limit in Eq.~(\ref{coding}) is required by possible superadditivity of the function
$C_{\chi}(\Phi^{\otimes n})$ as a consequence of  using entangled quantum  block-letter  $\rho^{(w)}$ at the encoding stage.

In the case where the property of {\it additivity} holds, i.e.
\begin{eqnarray}
C_{\chi }(\Phi ^{\otimes n})=n\; C_{\chi }(\Phi )\;,  \label{addi}
\end{eqnarray}
the regularization~(\ref{coding})  is not needed and  the capacity admits a simple
single letter expression $C(\Phi )=C_{\chi}(\Phi)$, where
\begin{eqnarray}
C_{\chi }(\Phi )=\max_{p_{x},\rho _{x}}\left\{ S\left(
\sum_{x}p_{x}\Phi \left[ \rho _{x}\right] \right)
-\sum_{x}p_{x}S\left( \Phi \left[ \rho _{x} \right] \right) \right\}
,  \label{cchi1}
\end{eqnarray}
and the maximization is over all finite ensembles of states $\rho _{x}$ taken with probabilities $p_{x}$.
 The additivity~(\ref{addi}) means that using entangled
states at the input of the channel $\Phi$ does not increase the quantity
of transmitted classical information. The validity of the property~(\ref{addi}) was
established for a number of channels,  including the unital qubit
channels \cite{king},  the depolarizing channel \cite{kingdep}, the
erasure channel~\cite{erasure}, the purely lossy Bosonic channel~\cite{gio},  and the
whole class of entanglement-breaking channels~\cite{ebc}.
In all of these cases except the last one  the analytical solution of the maximization
problem for $C_{\chi}(\Phi)$ is possible.
\newline

{\bf Example:} For the depolarizing channel of Eq.~(\ref{dep}), the additivity~(\ref{addi})
combined with the high symmetry of the map $\Phi$ allows one to find its classical capacity~\cite{kingdep},
\begin{eqnarray}
\fl \qquad C(\Phi )=C_{\chi }(\Phi )=\log_2 d+\Big(1-p\frac{d-1}{d}\Big)\log_2 \Big(1-p
\frac{d-1}{d}\Big)+p\frac{d-1}{d}\log_2 \frac{p}{d}\; ,  \label{c1dep}
\end{eqnarray}
with the maximum attained on the ensemble of $d$ equiprobable orthogonal pure states.
\newline

  The need of the regularization limit in Eq.~(\ref{coding}) for a generic channel
has been debated at length, for a survey see e.g.~\cite{h2007,RUSKAI}.
The question, if there exist at all nonadditive quantum channels, turned out to be extremely difficult and remained
open for rather a long time.
The additivity  of $C_{\chi}$ plays an important role in quantum information and is linked to the additivity
of other quantum entropic quantities~\cite{mats,Shb} including
the  minimal output entropy $\min_\rho S(\Phi[\rho])$.
A significant step forward was made by Shor~\cite{Shb}, who showed in particular that
proving the additivity of the minimal output entropy
for {\it all pairs} of channels,~i.e.
\begin{eqnarray}
\min_\rho S(\Phi_1\otimes \Phi_2[\rho])=
\min_\rho S(\Phi_1[\rho]) + \min_\rho S( \Phi_2[\rho]) \;, \qquad  \forall \Phi_1, \Phi_2\;, \label{minentr}
\end{eqnarray}
 is equivalent to proving the similar additivity for $C_{\chi}$, i.e.
\begin{eqnarray} \label{minchi}
C_\chi(\Phi_1\otimes \Phi_2)=
C_\chi(\Phi_1) + C_\chi( \Phi_2)\;, \qquad \forall \Phi_1, \Phi_2\;,
\end{eqnarray}
and hence it would imply (\ref{addi}). This in turn stimulated intensive research of several
related quantities, which besides the von Neumann entropy~\cite{kingruskai01},
include quantum R\'{e}nyi entropies
 see e.g. Refs.~\cite{AMOSOV1,KINGmax,knatr,sew,gio2005,alikifannes,DEVJKR}. 
 The possibility of violating the additivity for these functionals was
established in several cases~\cite{wernerholevo,wmult,hayden,hwint,Cubitt}, 
but the violations were not strong enough to imply superadditivity of $C_\chi$.
    The problem has been settled by
Hastings~\cite{hast} who proved that
channels which violate the additivity~(\ref{minentr}) and hence require
the regularization in Eq.~(\ref{coding})  do exist
 at least in very high dimensions (see \cite{Fuk} for
the actual dimensionality estimates),
  among mixtures of unitary channels of the form
 $\Phi \lbrack \rho] =\sum_{j}\pi _{j}U_{j}\rho U_{j}^{\dag }$
(here $\pi _{j}$ is a probability distribution while $U_{j}$ are unitary operators),
but so far no concrete example was found.
\newline 

 It is  finally worth mentioning that transmission of classical information through quantum  channels may display yet another form of superadditivity  of a
  higher complexity  level as compared to the one implied by the regularizations of Eqs.~(\ref{cc}) and (\ref{coding}).
 Namely, consider two different quantum channels $\Phi_1$ and $\Phi_2$  with the classical capacities $C(\Phi_1)$ and
 $C(\Phi_2)$ defined as in~(\ref{coding}). The problem is whether the
 capacity $C(\Phi_1 \otimes \Phi_2)$ of  the tensor product channel
  $\Phi_1 \otimes \Phi_2$ can be strictly greater than the sum of the two individual capacities, i.e.
 \begin{eqnarray}\label{openproblem}
 C(\Phi_1 \otimes \Phi_2) \stackrel{?}{>} C(\Phi_1) + C(\Phi_2)\;.
 \end{eqnarray}
It should be stressed that due to the regularization present in~(\ref{coding}) the superadditivity of the functional $C_\chi(\cdot)$ proved in Ref.~\cite{hast}
{\it is not} sufficient to answer this question. In fact, the strict inequality in~(\ref{openproblem})
would be similar to the superactivation property for the quantum capacities~\cite{SMITH}, see Sec.~\ref{sectionVAR}.
 Proving $C(\Phi_1 \otimes \Phi_2)\ge C(\Phi_1) + C(\Phi_2)$ is trivial: this follows simply from the possibility of operating the two channels independently, i.e. using codewords that factorize in the partition $\Phi_1\otimes \Phi_2$. However the possibility to have the strict inequality here
stems from the freedom of introducing the quantum correlations  at the input of such a partition.

\subsection{Entropy exchange and quantum mutual information}\label{entexch}

In the classical case correlation
between  two random
variables $X,Y$ (which, in particular, may describe input and output of a
classical channel) can be measured  by the mutual information $I(X;Y)$ introduced in Eq.~(\ref{DefSI}).
A quantum analog of this quantity has been identified  in Sec.~\ref{ccqc1} as the  $\chi$-information  of Eq.~(\ref{chiq}).
The quantity $\chi$ however does not account for all the correlations which can be established
between the input and the output of a quantum channel.
To characterize them one needs to look for other
quantum generalizations of $I(X;Y)$ (the possibility of
multiple quantum generalizations should not be surprising: ultimately
this is a consequence of the fact that  joint distribution of quantum observables
exists only in the very special case when they commute).

%%%%%%%%%%%%%%%%%%%%%%%%%%%%%%%%%%%%%%%%%%%%%%%%%%%%
\begin{figure}[t]
\begin{center}
\includegraphics[width=350pt]{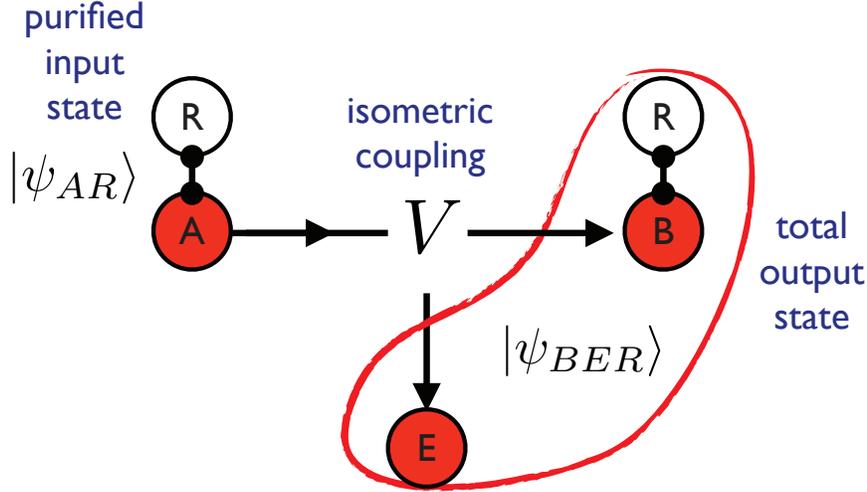}
\caption{Schematics of the purification of the information transfer through the
quantum channel~$\Phi$ connecting the input system $A$ to the output space $B$. Here
 $R$ is a reference system which is used to purify the input states $\rho_A$ of $A$.
Following the  analysis of Sec.~\ref{sec:ChannelsOpen}
the mapping $\Phi$ is then represented in terms of an isometric transformation $V$ that
maps vectors of $A$ into (possibly) entangled states of $B$ and of the environment $E$.
The reduced density matrix of $B$ and $E$ provide, respectively, the output states of the channel $\Phi$
 and of its complementary $\tilde{\Phi}$.
In the quantum wiretap model of Sec.~\ref{wiretap} the reduced density matrix of $E$ is assigned
to the eavesdropper.
}
\label{figu1}
\end{center}
\end{figure}
%%%%%%%%%%%%%%%%%%%%%%%%%%%%%%%%%%%%%%%%%%%%%%%%%%

One way to proceed is to exploit the following purification
trick~\cite{SCHUNIEL,sw97}. Consider a quantum channel  $\Phi =\Phi
_{A\rightarrow B}$ that maps the input states $\rho=\rho_A\in
\mathfrak{S} ({\cal H}_A)$ into the outputs $\rho_B= \Phi [\rho]$.
Let us introduce a reference system $ \mathcal{H}_{R}\simeq
\mathcal{H}_{A}$ and purify the input $\rho$ to $
\ketbra{\psi_{AR}}{\psi_{AR}}$ in the space $\mathcal{H}_{A}\otimes
\mathcal{H}_{R}.$ By Eq.~(\ref{eent}) the state $\rho _{R}$ of the
reference system has the same spectrum as $\rho$ and hence
$S(\rho)=S(\rho _{R}).$ In what follows, for the sake of simplicity,
we shall abbreviate the notations for the entropies of partial
states by omitting the symbol of the density operator $\rho ,$ so,
for example, the last equality will be written as $ S(A)=S(R).$ Now
let us focus on the entanglement transmission
scenario~(\ref{etrans}) where the state
$\ketbra{\psi_{AR}}{\psi_{AR}}$ is transmitted via the channel $\Phi
_{A\rightarrow B}\otimes \mathrm{Id}_{R}$, producing the output
state
\begin{eqnarray}
\rho _{BR}=(\Phi _{A\rightarrow B}\otimes
\mathrm{Id}_{R})[\ketbra{\psi_{AR}}{\psi_{AR}}]\;.
\end{eqnarray}
Consider the quantity
\begin{eqnarray}\label{qmu}
I(\rho ,\Phi )\equiv S(R)+S(B)-S(BR)\;,
\end{eqnarray}
which is nonnegative and vanishes if and only if $\rho _{BR}=\rho
_{B}\otimes \rho _{R}$, see Sec.~\ref{relentropy}. Since the
reference system $R$ is a copy of the input $A$ which remains intact
in the course of the transmission, one can interpret $I(\rho ,\Phi )$
as another substitute for the Shannon mutual information between the input
and the output of the quantum channel. It is called the {\it quantum
mutual information} \cite{l4,ac}. It is combined of the three
entropies: $S(R)=S(\rho )$ -- the input entropy,
$S(B)=S(\Phi \lbrack \rho ])$ -- the output entropy, and the joint
entropy $S(BR)$ which deserves a closer look in the isometric
representation~(\ref{compl}) of the channel $\Phi$.

A useful expression for the quantity~(\ref{qmu}) is obtained by
introducing  the channel environment $E$ and the isometry
$V:\mathcal{H}_{A}\rightarrow \mathcal{H}_{B}\otimes
\mathcal{H}_{E}$ which represent the transformation $\Phi$ as detailed in Sec.~\ref{irreversibility}, see Fig.~\ref{figu1}. At the end of the
transmission we have the tripartite system $BER$ characterized by
the space $\mathcal{H}_{B}\otimes \mathcal{H}_{E}\otimes
\mathcal{H}_{R}.$ The total output state $\rho _{BER}$ in
$\mathcal{H} _{B}\otimes \mathcal{H}_{E}\otimes \mathcal{H}_{R}$ is
pure and is described by the vector $ |\psi _{BER}\rangle =(V\otimes
I_{R})|\psi _{AR}\rangle .$ Looking at the split $BR|E$ we have
a pure state bipartite system with partial states $\rho
_{BR}=\mathrm{Tr}_{E}\rho _{BER}$, $\rho _{E}=\mathrm{Tr}_{BR}\rho
_{BER}$ whose entropies $S(BR)$ and $S(E)$ are equal, see
Eq.~(\ref{eent}). This implies
\begin{eqnarray}
S({BR})=S({E})\equiv S(\rho ,\Phi )\;,  \label{eepur}
\end{eqnarray}
where the last quantity is called the {\it entropy
exchange}~\cite{l4,SCHUNIEL,sw97}, as it measures the entropy change in the
environment (recall that in the isometric representation the initial
environment state is assumed pure). Notice also that by construction
$\rho_E$ is the state at the output of the complementary channel $\tilde{\Phi}$.
Therefore by Eq.~(\ref{cmpm}),
\begin{eqnarray}\label{enex}
S(\rho ,\Phi )=S(\tilde{\Phi}[\rho ])=S\left( \left[ \mathrm{Tr}\rho V_{\beta }^{\dag
}V_{\alpha } \right] _{\alpha ,\beta =1,\dots ,d_E}\right)\;.
\end{eqnarray}
The quantum mutual information can then be expressed as
\begin{eqnarray}
I(\rho ,\Phi )=S(\rho )+S(\Phi \lbrack \rho ])-S(\rho ,\Phi )\;.
\label{qme}
\end{eqnarray}
\newline

{\bf Example:} Consider the depolarizing channel (\ref{dep}) and the chaotic state
$I/d$.
By using the Kraus decomposition (\ref{depkraus}) we find that the state $\tilde{\Phi}[I/d]$
 has  eigenvalues 0 and
$1-p\frac{d^{2}-1}{d^{2}}$ of multiplicity 1, and  $\frac{p}{d^{2}}$ of
multiplicity $d^{2}-1$ (the zero appearing because of the linear
dependence of the Kraus operators). Hence,
\begin{eqnarray}
\fl \qquad S(I/d,\Phi )=-\Big(1-p\frac{d^{2}-1}{d^{2}}\Big)\log_2
\Big(1-p\frac{ d^{2}-1}{d^{2}}\Big)-p\frac{d^{2}-1}{d^{2}}\log_2
\frac{p}{d^{2}}\;. \label{ExEn}
\end{eqnarray}
Combining with the input and the output entropies $S(I/d)=S(\Phi
\lbrack I/d])=\log_2 d,$ we obtain
\begin{eqnarray}
\fl \qquad I(I/d,\Phi )=\log_2 d^{2}+\Big(1-p\frac{d^{2}-1}{d^{2}}\Big)\log_2 \Big(
1-p\frac{d^{2}-1}{d^{2}}\Big)+p\frac{d^{2}-1}{d^{2}}\log_2 \frac{p}{d^{2}}\;.
\label{qme_dep}
\end{eqnarray}
\newline

Up to now we have encountered three entropic quantities: the input
entropy $S(\rho ),$ the output entropy $S(\Phi \lbrack \rho ])$ and
the entropy exchange $S(\rho ,\Phi ).$ By making different bipartite
splits of the system $BER,$ we obtain similarly to (\ref{eepur}) the
identities
\begin{eqnarray}
S({BE}) &=&S({R})\equiv S(\rho )\;,  \label{dop1} \\
S({RE}) &=&S({B})\equiv S(\Phi \lbrack \rho ])\;,
\label{dop2}
\end{eqnarray}
and another two information quantities: the {\it loss}, i.e. the
quantum mutual information between the input and the environment:
\begin{eqnarray}
L(\rho ,\Phi ) \equiv S(R)+S(E)-S(RE)=S(\rho )+S(\rho ,\Phi )-S(\Phi
\lbrack \rho ])\;,  \label{loss}
\end{eqnarray}
and the {\it noise}, i.e. the quantum mutual information between
 the output and the environment
\begin{eqnarray}
N(\rho ,\Phi ) \equiv S(E)+S(B)-S(EB)=S(\rho ,\Phi )+S(\Phi \lbrack \rho
])-S(\rho )\;.
\end{eqnarray}
The nonnegativity of the quantities $I(\rho ,\Phi ),$ $L(\rho ,\Phi )$
and $ N(\rho ,\Phi )$ along with (\ref{dop1}), (\ref{dop2}) implies
that the basic entropies $S(\rho )$, $S(\Phi \lbrack \rho ])$, and $S(\rho,\Phi )$ satisfy all the triangle inequalities, namely,
\begin{eqnarray}
\left\vert S(\Phi \lbrack \rho ])-S(\rho ,\Phi )\right\vert \leq S(\rho )\;,
\end{eqnarray}
and two other inequalities obtained by cyclic permutations.

As in the case of the Shannon mutual information~(\ref{DefSI}),
the quantities $I(\rho ,\Phi ),$ $L(\rho ,\Phi )$ and $N(\rho ,\Phi)$  can also be expressed in terms of conditional (quantum) entropies. For instance we have
\begin{eqnarray}
\fl \quad  I(\rho ,\Phi )=S(B)-S(B|R)\;,\qquad S(B|R)\equiv S(BR)-S(R)=S(\rho ,\Phi)-S(\rho)\;,
\end{eqnarray}
with $S(B|R)$ being the quantum conditional entropy of $B$ given $R$. Notice however that while
in the classical case the conditional entropy is always non-negative, in the quantum scenario
this is no longer valid as the quantum entropy is not necessarily monotone
with respect to enlargement of the system, i.e.  $S(R)\not\leq
S(BR)$ (an extreme example of this property is obtained
 when $BR$ is a pure entangled system:
in this case $S(BR)=0$ while $S(R)>0$ -- this cannot   happen in the classical statistics where the
partial state of a pure state can never be mixed, see also Sec.~\ref{sec:ClassicalQuantum}).
An operational interpretation of the possible negativity of
the conditional entropy in terms of ``quantum state merging'' is given in Ref.~\cite{wo}.
Quite remarkably, the monotonicity of the
conditional entropy still holds:  for any composite system $ABC$ one has
\begin{eqnarray}
S(A|BC)\leq S(A|B)\;,  \label{monce}
\end{eqnarray}
and it is this property which makes the quantum conditional entropy
useful. When written in terms of (unconditional) quantum entropy,
the inequality~(\ref{monce}) amounts to the fundamental property of
{\it strong subadditivity}:
\begin{eqnarray}
S({ABC})+S({B})\leq S({AB})+S({BC})\;.
\label{ssa}
\end{eqnarray}
The proof of this
very useful inequality was  given first by Lieb and Ruskai in~Ref.~\cite{lieb} and  remains rather
involved in the quantum case even after subsequent simplifications,  see e.g.~\cite{RUSKAIPROOF}.

Due to the strong subadditivity (\ref{ssa}), the quantum mutual
information enjoys important properties similar to that of the
Shannon information in the classical case~\cite{ac,nc}. Specifically, given two quantum channels $\Phi_1$ and $\Phi_2$, one has
\begin{eqnarray} \label{subqminfo}
\fl \mbox{(i) subadditivity:}  \qquad && I(\rho _{12},\Phi _{1}\otimes \Phi _{2})\leq I(\rho
_{1},\Phi _{1})+I(\rho _{2},\Phi _{2})\;; \\
\fl \mbox{(ii) data processing inequalities:}
\qquad && I(\rho ,\Phi _{2}\circ \Phi _{1})\leq \min \{I(\rho ,\Phi _{1}),I(\Phi
_{1}(\rho ),\Phi _{2})\}\;. \label{subqminfo1}
\end{eqnarray}

\subsection{Entanglement as an information resource}

 In the previous sections we have seen that entangled decodings
can enhance transmission of classical information through quantum
channel. An even more impressive gain is achieved when entanglement
between the input and the output of the channel is available. The
classical capacity of a memoryless communication line defined by a quantum channel $\Phi $ can be
substantially increased by using this additional resource in spite
of the fact that entanglement alone cannot be used to transmit
information. This fundamental observation was first made by
Bennett and Wiesner who introduced the notion of {\it superdense coding}~\cite{wisner},
see also~\cite{BAN,BRAUKI,RALPHHUNT} for generalizations  to continuous
variables and~\cite{MATTLE,PEREIRA,LI,MIZUNO} for experimental tests.
Here as in some other cases entanglement plays a role
of ``catalyser'', disclosing latent information resources of a
quantum system.

The scenario of {\it entanglement-assisted} communication assumes that
prior to the information transmission parts of an entangled state
$\rho _{AB}$ are distributed between sender $A$ and receiver $B.$
Then sender $A$ encodes the classical messages $w$ into different
operations $\mathcal{E}_{w}$ on his part of the entangled state, and
the result of these operations is sent to $ B $ via the quantum
channel $\Phi .$ Thus at the end of the transmission the state $\left(
\Phi \circ \mathcal{E}_{w}\otimes \mathrm{Id}_{B}\right) \left[ \rho
_{AB}\right] $ is available to the receiver which extracts the
classical information by making quantum measurement on this state. Optimizing
the transmission rate with respect to the entangled states $\rho
_{AB},$ (block) encodings of $A$ and measurements (decodings) of $B$
gives the {\it classical entanglement-assisted capacity}. The
corresponding {\it coding theorem} of Bennett, Shor, Smolin and
Thapliyal provides a simple formula (the proof of which is far from
being simple,~\cite{bsst2,bsst,holea}) giving an operational
characterization for the
 quantum mutual information (\ref{qme}):
\begin{eqnarray}
C_{ea}(\Phi )=\max_{\rho }I(\rho ,\Phi )\;,  \label{ceai}
\end{eqnarray}
where the maximum is taken over all possible input states $\rho$ of a single channel use.
(Equation~(\ref{ceai}) refers to the case of unlimited shared entanglement. For analysis of
the case in which only limited resources are available see Refs.~\cite{Shb,BOWEN,DHW}).
Remarkably, unlike Eq.~(\ref{coding}), the capacity formula~(\ref{ceai}) does
not contain the limit $n\rightarrow \infty$
because $C_{ea}(\Phi^{\otimes n})=n\; C_{ea}(\Phi )$ due to subadditivity~(\ref{subqminfo}) of the quantum
mutual information. Also by construction it follows that entanglement-assisted capacity  $C_{ea}(\Phi)$ is always
greater than or equal to its unassisted counterpart $C(\Phi)$.

If $\Phi =\mathrm{Id}$ is the noiseless channel then this scenario
coincides with the original superdense coding protocol~\cite{wisner} for which
$C_{ea}(\mathrm{Id})=2\log_2 d=2C(\mathrm{Id})$, so the
capacity gain is equal to 2. Similar doubling of the capacity holds
for the quantum erasure channel. However in general, the more noisy
is the channel  -- the greater is the gain, and in the limit of very
noisy channels the gain can be arbitrarily large. For example, in
the case of depolarizing channel (\ref{dep}), the maximum in
(\ref{ceai}) is attained on the chaotic state so that $C_{ea}(\Phi
)$ is given by the formula (\ref {qme_dep}). Comparing this with the
quantity $C(\Phi )$ given by the formula (\ref{c1dep}), one sees
that the gain $\frac{C_{ea}(\Phi )}{C(\Phi )}\rightarrow d+1$ in the
limit of large noise $p\rightarrow 1$.

Interestingly, one can have $C_{ea}(\Phi)>C(\Phi)$ for entanglement-breaking
channel, for example, this holds for the depolarizing channel with
$p\ge d/(d+1)$. The explanation given in \cite{bsst} is that
unassisted classical data transmission through entanglement-breaking
channel involves communication cost which is always greater than the
difference between $C_{ea}(\Phi)$ and $C(\Phi)$.

A protocol in a sense dual to superdense coding is {\it quantum teleportation} which was first introduced in Ref.~\cite{telep}.
Quantum information theory predicts the possibility of a nontrivial way of transmitting arbitrary quantum state
$\rho$, when the state carrier is not transferred physically
but only some classical information is transmitted trough a classical communication line, see also Refs.~\cite{TELEBRAU,TELEMILBURN,TELEVAN} for the
generalization to continuous variables and Refs.~\cite{ZEIL,BOSCHI} for the first experimental test.
 The necessary
additional resource here is a shared entangled state
between the sender
and the receiver of the classical information (it is impossible to
reduce the transmission of an arbitrary quantum state solely to
sending the classical information: since the classical information
can be copied, it would mean the possibility of cloning the quantum
information~\cite{WZ}).
The  effective maps $\Phi$  resulting from such protocols are  called
{\it quantum teleportation channels} ~\cite{HTELEP,POPE,BOBO,wpgg}.
They  represent a proper subset of the class of quantum communication systems which
can be fully specified by assigning a joint initial state $\rho_{AB}$,
characterizing the shared entanglement between the sender and the receiver, and the
local operations the two parties are supposed to perform on it.

\subsection{Coherent information and perfect error correction}

An important part of the quantum mutual information $I(\rho ,\Phi )$
is the {\it coherent information}~\cite{SCHUNIEL}
\begin{eqnarray}
I_{c}(\rho ,\Phi ) &\equiv&S(\Phi \lbrack \rho ])-S(\rho, \Phi ) \nonumber \\
&=&S(B)-S(E) \nonumber \\
&=&S(B)-S(BR) \nonumber \\
&=&-S(R|B)\;. \label{coherent}
\end{eqnarray}
This quantity is closely related to the quantum capacity of the channel $\Phi$
which will be considered in the next Section.

The coherent information {\it does not} share some ``natural''
properties of quantum mutual information such as subadditivity and
the second data processing inequality. Moreover similarly to its
classical analog $H(Y)-H(XY)=-H(X|\,Y)$ it can be negative.
However $I_{c}(\rho ,\Phi )$ satisfies the
first data processing inequality:
\begin{eqnarray}
I_{c}(\rho ,\Phi _{2}\circ \Phi _{1})\leq I_{c}(\rho ,\Phi _{1})\;,
\label{eq28}
\end{eqnarray}
which follows from the relation $I_{c}(\rho ,\Phi )=I(\rho ,\Phi
)-S(\rho )$ and from the corresponding property of quantum mutual
information $I(\rho ,\Phi )$.

There is a close connection between perfect transmission of quantum
information, error correction and certain property of coherent
information. The channel $\Phi $ is {\it perfectly reversible on
the state} $\rho =\rho _{A}$, if there exists a {\it recovery
channel} $\mathcal{D}$ from $B$ to $ A$, such that
\begin{eqnarray}
(\mathcal{D}\circ \Phi \otimes \mathrm{Id}_{R})[\rho _{AR}]=\rho _{AR}\;,
\end{eqnarray}
where $\rho _{AR}$ is a purification of the state $\rho _{A}$ with the
reference system $R.$

One can show \cite{bns} that the following statements are
equivalent:

\begin{enumerate}
\item the channel $\Phi$ is perfectly reversible on the state $\rho$;
\item $L(\rho ,\Phi )=0$, that is $\rho _{RE}=\rho _{R}\otimes \rho _{E}$;
\item $I_{c}(\rho ,\Phi )=S(\rho ).$
\end{enumerate}
The condition (ii) means that information does not leak into the
environment, i.e. the channel is ``secret'' or ``private''. Thus
the perfect reversibility of the channel is equivalent to its privacy.
The condition (iii) means that under private transmission of the
state $\rho $ by the channel $\Phi $, the coherent information
$I_{c}(\rho ,\Phi )$ should attain its maximal value $ S(\rho )$.
The equivalence of  (ii) and (iii) is obvious since $S(\rho
)-I_{c}(\rho ,\Phi )=S(\rho )+S(\rho ;\Phi )-S(\Phi \lbrack \rho
])=L(\rho ,\Phi )\geq 0$ by Eq.~(\ref{loss}).

In particular, choosing the chaotic state $\rho=I_{A}/d_{A}$,
we get the condition $\log_2 d_{A}=I_{c}(I_{A}/d_{A},\Phi )$. This shows
that the coherent information should be related to the quantum capacity of
the channel $\Phi $, which characterizes the maximal dimension of the
perfectly transmittable states. (In fact, such a relation is valid for
asymptotically perfect transmission through the block channel $\Phi
^{\otimes n}$ when $n\rightarrow \infty ,$ as we shall see in the following
Section). In support to this argument we notice here that the perfect reversibility property of
$\Phi$ on  $\rho$ can also be stated by
saying that there exists a recovery channel ${\cal D}$ such that
\begin{eqnarray}
\mathcal{D}\circ \Phi \lbrack {\rho}']={\rho}'\;,
\end{eqnarray}
for all states ${\rho}'$ with $\mbox{supp}[{\rho}']\subset
\mathcal{L} \equiv \mbox{supp}[\rho]$, with $\mbox{supp}[\rho]$
being the support of the state~$\rho$. We can express the same
property by saying that $\Phi $ is {\it perfectly reversible on the
subspace} $\mathcal{L}$. In other words, the subspace
 $\mathcal{L}$
  is a quantum code correcting the error described by the noisy channel $\Phi$
  (and hence, all the related errors $\rho \rightarrow V_{j}\rho V_{j}^{\dag },$ where
$\Phi \left[ \rho \right] =\sum_{j}V_{j}\rho V_{j}^{\dag }$)~\cite{kl}.

On the other hand, notice that by the formula (\ref{eepur}) $S(\rho
;\Phi )=S(\tilde{\Phi}[\rho ]),$ where $\tilde{\Phi}$ is the
complementary channel, hence
\begin{eqnarray}
I_{c}(\rho ,\Phi )=S(\Phi \lbrack \rho ])-S(\tilde{\Phi}[\rho
])=-I_{c}(\rho ,\tilde{\Phi}).\label{complcoher}
\end{eqnarray}
The next statement which generalizes observation at the end of Sec.~\ref{complement}
gives a characterization in terms of
complementary channel which underlies the coding theorem for the
secret classical capacity of the channel in Sec. \ref{pereh}. The
following conditions are equivalent:

\begin{enumerate}
\item the channel $\Phi$ is perfectly reversible on the subspace $\mathcal{L}$;
\item the complementary channel (\ref{cmpm}) is completely
depolarizing on $\mathcal{L}$, i.e.
\begin{eqnarray}
\tilde{\Phi}[{\rho'}]=\rho _{E}\;,  \label{depL}
\end{eqnarray}
for any state ${\rho}'$ with $\mbox{supp}[{\rho}']\subset
\mathcal{L}$, where $\rho_{E}$ is a fixed state.
\end{enumerate}

\subsection{The quantum capacity}\label{qcap}

The transformation $\rho \rightarrow \Phi [\rho ]$ of quantum states can be regarded as the transfer of
quantum information.
The discovery of the {\it quantum error-correcting codes}
\cite{ecc,s} is related to the question of asymptotically (as $n\rightarrow
\infty $) error-free transmission of quantum information by the
channel $\Phi ^{\otimes n}$. The {\it quantum
capacity} $Q(\Phi)$ is defined as the maximum amount of quantum
information per one use of the channel which can be transmitted with asymptotically vanishing
error~\cite{bs,ll,bkn,dev,kretch}. It is related to the
dimensionality of the subspace of state vectors in the input space
($\approx 2^{nQ(\Phi )}$) that are transmitted asymptotically
error-free. For the quantum capacity there is an expression in terms
of coherent information~(\ref{coherent}), namely
\begin{eqnarray}
Q(\Phi)=\lim_{n\rightarrow \infty }\frac{1}{n}\max_{\rho ^{(n)}}I_{c}(\rho
^{(n)},\Phi ^{\otimes n})\;,  \label{UBC}
\end{eqnarray}
the maximum being performed over all input states $\rho^{(n)}$ of $n$ successive channel uses.
The relation between $Q(\Phi)$ and the coherent information of the channel was conjectured in Ref.~\cite{ll}
and made more precise in Refs.~\cite{bns,shorproof}, while the ultimate proof of Eq.~(\ref{UBC})
was given by Devetak~\cite{dev} exploiting the fact
that the quantum capacity of a channel is closely related to its
cryptographic characteristics, such as the capacity for
the secret transmission of classical information and the rate of the
random key distribution.
More specifically, as discussed in detail in the following Section, a deep analogy
with a {\it  wiretap channel}~\cite{WYNER} was used, the
role of the eavesdropper in the quantum case played by the environment
of the system.

For the ideal  channel $\mathrm{Id}$  one easily verifies that $Q(\mathrm{Id})=\log_2 d$.
Analytical expression for the capacity of quantum depolarizing
channel~(\ref{dep}) is still unknown in the general case, although there are
fairly close lower~\cite{eof,vss,HAMADA} and upper~\cite{VED,RAINS1,RAINS2,SSW} bounds for it. A major
difficulty of evaluating $Q(\Phi)$ lies in the nonadditivity of the
quantity $I_{c}(\rho ^{(n)},\Phi ^{\otimes n})$ \cite{vss}.
However, there is an important class of  degradable channels~\cite{ds,krs}
for which the additivity holds, so one can replace  Eq.~(\ref{UBC})  with
the  convenient  ``one-letter'' formula
\begin{eqnarray}
Q(\Phi) = Q_{1}(\Phi )\equiv \max_{\rho }I_{c}\left( \rho ,\Phi \right) \;,  \label{Q1}
\end{eqnarray}
where now the maximization is taken over the input states $\rho$
for single channel use. The channel $\Phi$ is called {\it degradable} if there
exists  quantum channel $\Upsilon$ such that
 $\tilde{\Phi}=\Upsilon\circ \Phi$~\cite{ds}. In practical terms, this relation expresses the fact that
 the complementary channel $\tilde{\Phi}$ is ``more noisy'' than $\Phi$ (it can be obtained
 from the latter by ``adding'' the extra noise~$\Upsilon$).
Notable examples of degradable channels are the dephasing
channels~\cite{ds,mr} and the amplitude damping
channel~\cite{giofazio} for which the maximization in~(\ref{Q1}) can
be explicitly performed. Another class of interest is constituted by
the so called~{\it anti-degradable} channels: $\Phi$ is called
{anti-degradable}~\cite{cgh,giofazio}, if there exist a channel
$\Upsilon^{\prime},$ such that $\Phi =\Upsilon^{\prime }\circ
\tilde{\Phi}.$ Apparently $\Phi $ is degradable if and only if
$\tilde{\Phi}$ is anti-degradable. Any  anti-degradable channel $\Phi$
has the null quantum capacity,
$Q(\Phi )=0$. A formal proof of this fact will be given in Sec.
\ref{relentropy}; there is  however a simple  heuristic argument
based on  no-cloning theorem, see Eq.~(\ref{cloning}), that
explains why $Q(\Phi)$ cannot be positive. Suppose that
$Q(\Phi)>0$ for anti-degradable channel $\Phi$. This implies that, via
encoding and decodings protocols, the sender will be able to
transfer to the receiver at the output of the channel $\Phi$ an
arbitrary (unknown) pure quantum state $|\psi\rangle$. Since $\Phi$ is
anti-degradable, this implies that the same protocol will also
allow to recover the same quantum message at the output of the
complementary channel $\tilde{\Phi}$ (one can
reconstruct the associated output of $\Phi$ by applying the channel
$\Upsilon'$). Now the contradiction arises by observing that  in the
isometric representation introduced in Sec.~\ref{entexch}, $\Phi$ and
$\tilde{\Phi}$ describe the two reduced quantum information flows
that enter, respectively, to the receiver  and to the channel
environment. Two independent observers collecting those data will
thus be able to get a copy of the same state $|\psi\rangle$,
realizing {\it de facto} a cloning machine, which is impossible.
Notable examples of anti-degradable channels are provided by the
entanglement-breaking channel (\ref{eb}) whose   complementary
counterparts are the dephasing channel (\ref{gdiag}) (the channel
$\Upsilon'$ in this case can be taken as $\Upsilon^{\prime
}[\rho ]=\sum_{\alpha} |\varphi _{\alpha }\rangle \langle e_{\alpha
}|\rho |e_{\alpha }\rangle \langle \varphi _{\alpha }|$).
\newline

{\bf Example:} The quantum erasure channel $\Phi _{p}$ introduced in Eq.~(\ref{erasure})
is degradable for $p\in \lbrack 0,1/2]$ and
anti-degradable for $p\in \lbrack 1/2,1]$. Its quantum capacity is computed~\cite{erasure} as
\begin{eqnarray}
Q(\Phi _{p})=\left\{
\begin{array}{ll}
(1-2p)\log_2 d\;,\quad & p\in \lbrack 0,1/2]\;; \\
0\;, & p\in \lbrack 1/2,1]\;.
\end{array}
\right.  \label{qerasure}
\end{eqnarray}
\newline

Another  class of channels which, as the anti-degradable ones,  posses null quantum capacity is given by
the so called   PPT ({\it partial positive transpose}) or {\it binding} channels~\cite{HBIND}.
These  maps  include as a special case the entanglement-breaking channels of Sec.~\ref{q-c-q}, and are characterized by the property that their associated  Choi-Jamiolkowski
state is PPT, i.e. it remains positive under partial transposition in the reference space. States that possess this property might not be separable but  their  entanglement (called {\it bound entanglement})
is ``weak" as it doesn't allow for {\it distillation}~\cite{HBOUND,HHHH}.
Binding  maps (if not entanglement-breaking) allow for a certain amount of entanglement transfer: the latter however is always non-distillable and, even though it admits private communication between the sender and the receiver~(see Sec.~\ref{wiretap}), it cannot be used for the faithful transfer of quantum information (not even in the presence of a two-way classical communication side line)~\cite{HBIND}.
For a recent study of the connections between anti-degradable and PPT channels, as well as on the characterization of the set of maps that have null quantum capacity see Ref.~\cite{SSINC}.

\subsection{Quantum wiretap channel} \label{wiretap}

\label{pereh}

Here we review in brief the argument behind the derivation of
Eq.~(\ref{UBC}). Consider the situation of classical information
transmission in which there are three parties: sender $A$, receiver
$B$ and the eavesdropper $E.$ A mathematical model of the
{\it quantum wiretap channel} comprises three Hilbert spaces
$\mathcal{H}_{A},\mathcal{H}_{B},\mathcal{H}_{E}$ and the isometric
map $V:\mathcal{H}_{A}\rightarrow \mathcal{H}_{B}\otimes \mathcal{H
}_{E},$ which transforms the input state $\rho _{A}$ into the state
$\rho _{BE}=\Phi _{A\rightarrow BE}\left[ \rho _{A}\right] \equiv
V\rho _{A}V^{\dag }$ of the system $BE,$ with partial states
\begin{eqnarray}
\rho _{B}=\Phi \left[ \rho _{A}\right] \equiv \mathrm{Tr}_{E}V\rho
_{A}V^{\dag },\qquad \quad  \rho _{E}=\tilde{\Phi} \left[ \rho _{A}\right] \equiv
\mathrm{Tr}_{B}V\rho _{A}V^{\dag }\;.
\end{eqnarray}
Notice that this description is formally identical to that of the complementary channels in Sec. \ref{complement}, where $E$ denoted the environment (in the wiretap model this is supposed to be
completely under control of the eavesdropper).

Assume now that $A$
sends the states $\left\{ \rho _{A}^{x}\right\} $ with probabilities $\{p_{x}\};$ then the parties $B$ and $E$ receive, correspondingly, the states $
\left\{ \rho _{B}^{x}\right\} $ and $\left\{ \rho _{E}^{x}\right\},$ and
upper bounds for Shannon informations they receive are the quantities
$\chi \left( \{p _{x}\},\{ \rho_{B}^{x}\}\right) $ and $\chi \left( \{p
_{x}\},\{ \rho _{E}^{x}\}\right)$, as stated in Eq.~(\ref{ebound}). In analogy with the classical wiretap
channel~\cite{WYNER} the ``secrecy'' of the transmission can be characterized by  the quantity $\chi \left( \{p _{x}\},\{\rho _{B}^{x}\}\right) -\chi \left(
\{p _{x}\},\{\rho _{E}^{x}\}\right)$ (here by secrecy we mean the amount of information which
can be shared between the $A$ and $B$ without  informing $E$). In fact, the capacity for secret transmission of classical information
is shown to be~\cite{dev}
\begin{eqnarray}
\fl \qquad C_{p}\left( \Phi _{A\rightarrow BE}\right) =\lim_{n\rightarrow \infty }\frac{1}{n}
\max_{p^{(n)},\Sigma ^{(n)}}
\left[ \chi \left( \{p_{i}^{(n)}\},\{\rho
_{B^{(n)}}^{i}\}\right) -\chi \left( \{p _{i}^{(n)}\},\{ \rho
_{E^{(n)}}^{i}\}\right) \right] ,  \label{secret}
\end{eqnarray}
where the maximum is taken over the families of states $\Sigma
^{(n)}\equiv\left\{ \rho _{A^{(n)}}^{i}\right\} $ in $\mathcal{H}_{A}^{\otimes n}$
and the probability distributions $p ^{(n)}\equiv \{p _{i}^{(n)}\}$ (we use
the notations $\rho _{B^{(n)}}^{i}=\Phi ^{\otimes n}\left[ \rho
_{A^{(n)}}^{i}\right] ,\rho _{E^{(n)}}^{i}=\tilde{\Phi} ^{\otimes n}\left[ \rho
_{A^{(n)}}^{i}\right] $).

Assuming that the input states $\rho _{A}^{x}$ are {\it pure}, and
denoting by $\rho\equiv {\rho}_{A}=\sum_{x}p_{x}\rho _{A}^{x}$ the
average state of the input ensemble, from (\ref{key2}) we obtain the
key relation~\cite{sw}
\begin{eqnarray}
I_{c}(\rho,\Phi ) &=&S(\Phi \lbrack \rho ])-S(\tilde{\Phi}[\rho ])
\nonumber
\\
&=&\left[ S(\Phi \lbrack \rho ])-\sum_{x}p _{x}S(\Phi \lbrack \rho
_{A}^x])
\right] -\left[ S(\tilde{\Phi}[\rho ])-\sum_{x}p _{x}S(\tilde{\Phi}[\rho
_{A}^x])
\right]  \label{key1} \nonumber \\
&=&\chi \left( \{p _{x}\},\{\Phi \lbrack \rho _{A}^x]\}\right) -\chi \left(
\{p _{x}\},\{\tilde{\Phi}[\rho _{A}^x]\}\right)\;, \label{key2}
\end{eqnarray}
where we used  Eq.~(\ref{complcoher}) and  the fact that the states $\rho _{BE}^x=V\rho
_{A}^xV^{\dag }$ are pure implying
\begin{eqnarray}
S(\Phi \lbrack \rho_{A}^x])=S\left( \rho_{B}^x\right) =S\left( \rho
_{E}^x\right) =S(\tilde{\Phi}[\rho_{A}^x])\;,
\end{eqnarray}
for all $x$.
The identity  (\ref{key2})
provides the fundamental connection
 between the quantum and the secret classical capacities
 which underlies the proof of Eq.~(\ref{UBC}) given in Ref.~\cite{dev}.
 Here we only notice that since
 in the computation of $C_{p}\left( \Phi _{A\rightarrow BE}\right)$ one takes into account
 all ensembles (not just the pure ones for which Eq.~(\ref{key2}) hold), we get the following inequality
\begin{eqnarray}
C_{p}\left( \Phi _{A\rightarrow BE}\right) \geq Q\left( \Phi \right)  \;,
\end{eqnarray}
which in general  can be strict -- see e.g. Refs.~\cite{HHHO,HHHLO} -- with the notable exception of
 degradable channels for which
\begin{eqnarray}  \label{QCp}
C_{p}\left( \Phi _{A\rightarrow BE}\right)=Q\left( \Phi \right)=Q_1\left( \Phi
\right).
\end{eqnarray}

To conclude this discussion of the wiretap channels we briefly mention the vast field of quantum cryptography which constitutes a self-consistent portion of quantum
information science  (for reviews relating this subject we refer to~\cite{GISIN,WEEDBROOK}).

\subsection{Capacities for Gaussian channels}\label{gauscapacity}
In this section  we briefly discuss the classical and the quantum capacities for Gaussian channels: in particular we focus on the
single-mode CPTP maps analyzed in Sec.~\ref{onemode}.

When speaking of the classical capacities for continuous-variables systems, to obtain reasonable finite quantities,
one should introduce an energy constraint onto input ensembles
  similarly to what was done for c-q Gaussian
channel in Example~3 of Sec.~\ref{ccqc1}. On the other hand, for the quantum capacity,  even though constraining the input may be reasonable from a practical point of view,
this is not strictly necessary as $Q(\Phi)$
remains finite even in the unbounded case.
Therefore in what follows we define $C_{\chi }(\Phi ,E)$
to be the value of~(\ref{cchi1}) where the maximum is taken
over ensembles $\left\{ p_x,\rho_x\right\} $ with the average
state ${\rho}=\sum_x p_x \rho_x$ satisfying the energy constraint
\begin{eqnarray}\label{maxen}
\mathrm{Tr} {\rho}a^{\dag }a\leq E\;,
\end{eqnarray}
($a$ and $a^\dag$ being the annihilation and creation operator of the mode).
Analogously, the entanglement-assisted classical capacity $C_{ea}(\Phi, E)$ is defined by the expression (\ref{ceai})
where the maximum taken over the states $\rho$ satisfying Eq.~(\ref{maxen}).

It is not known in general if $C_{\chi}(\Phi ,E)$ is additive under taking
tensor products of Gaussian channels which prevents from identifying $
C_{\chi }(\Phi ,E)$ with the full constrained capacity $C(\Phi ,E).$
However the additivity property (\ref{addi}) holds for entanglement-breaking
channels satisfying (\ref{doma}). Another case where additivity of $C_{\chi
}(\Phi ,E)$ with the energy constraint was established is the case of pure
loss channel, i.e. attenuator ($k<1$) with $N_0=0$ (environment in the vacuum
state) \cite{gio}. The actual computation of $C_{\chi}(\Phi ,E)$ is in
general also an open problem: there is a natural conjecture that the
maximum in $C_{\chi }(\Phi ,E)$ for a quantum Gaussian channel with
quadratic energy constraint is attained on a Gaussian ensemble of pure
Gaussian states~\cite{hw,SHAPIRO,wgc}, but so far this was only established for c-q channels and
the pure loss channel \cite{gio}. It is also worth pointing out that this conjecture
can be conveniently reformulated in terms of a property of the minimum output entropy~\cite{gmin,cong,RAUL}.
If the conjecture is true, then in the cases (i), (ii), and (iii) defined in Sec.~\ref{onemode} the optimal ensemble is the continuous
 ensemble formed by the coherent states  $|\zeta\rangle\langle \zeta|$
 distributed with a Gaussian density $p(\zeta)=\frac{1}{\pi E}
\exp \left( -{|\zeta|^{2}}/{E}\right)$, yielding
\begin{eqnarray}
C(\Phi ,E)=g(N'(E))-g(N'(0))\;,
\end{eqnarray}
where $N'(E)=k^{2}E+\max\{0,k^{2}-1\} +N_0$ is the mean number of quanta at the output when the input number of quanta is $E$, and where
 $g(x)$ is defined as in Eq.~(\ref{defig}).

The computation of the entanglement-assisted capacity $C_{ea}(\Phi,E)$ is a relatively simpler problem as, on one hand, no regularization over multiple
uses is required, and on the other hand, the quantity to be maximized (the quantum mutual information) is a concave function~\cite{hw}
which, even in the presence of  the  linear constraint~(\ref{maxen}), admits  a regular method of solution. In particular, for a Gaussian channel the maximum is always attained
on a Gaussian state which can be found as a solution of certain Kuhn-Tucker equations.
The expression of $C_{ea}(\Phi,E)$ for the one-mode channels was derived in Ref.~\cite{hw}  and generalized to the multimode case in Ref.~\cite{glmspra2003}.
Explicitly it is given by
\begin{eqnarray}
C_{ea}(\Phi ,E)=g(E) +g(N'(E)) -g(D_+(E)/2) -g(D_-(E)/2) \;,
\end{eqnarray}
where $g(x)$ and $N'(E)$ as before and where
 $D_{\pm}(E)\equiv  D(E)+N'(E)\pm E-1$, with $D(E)\equiv \sqrt{(E+N'(E)+1)^2 - 4 k^2 E (E+1)}$.
\newline

In general entanglement-breaking channels have zero quantum capacity $Q(\Phi)=0$.
However in any case the domain (\ref{doma}) is superseded by the
broader domain $N_0\geq \min \left( 1,k^{2}\right) -1/2$ where the channel is
anti-degradable and hence has zero quantum capacity~\cite{cgh}. On the other
hand, in the case $N_{0}=0$, $k^{2}>1/2$ the channel is degradable~\cite{cg,cgh},
hence the quantum capacity of the attenuation/amplification channel with $
N_{0}=0$ and the coefficient $k$ is equal to the maximized single-letter
Gaussian coherent information
\begin{eqnarray}\label{qcapgaussian}
Q(\Phi )=\sup_{\rho} I_c(\rho ,\Phi )=\max \left\{0,\log_2 \frac{k^{2}}{|k^{2}-1|}\right\}\;,
\end{eqnarray}
an expression conjectured in Ref.~\cite{hw} and proved in Ref.~\cite{wpgg}.
The case with $N_{0}>0$ remains open
question. Enforcing the energy constraint as in Eq.~(\ref{maxen}),  the above expression is replaced by
\begin{eqnarray}\label{qcapgaussian1}
Q(\Phi,E)=\max \left\{0, g(N'(E)) -g(D_+(E)/2) -g(D_-(E)/2) \right\}\;.
\end{eqnarray}

\begin{figure}[t]
\begin{center}
\includegraphics[width=330pt]{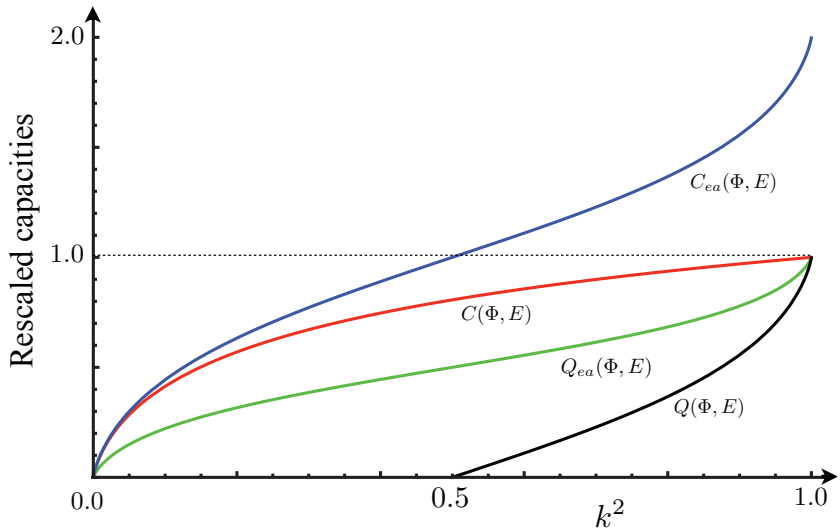}
\caption{(Color online)  Capacities~(\ref{capplot})
of the attenuation Bosonic Gaussian channel~(\ref{param}) for $N_0=0$ as a function of the attenuation parameter $k^2$ and for an energy constraint~(\ref{maxen})
with $E=10$. Specifically: classical capacity $C(\Phi,E)$~\cite{gio}(red curve); quantum capacity $Q(\Phi,E)$~\cite{wpgg} (black curve);
entanglement-assisted classical capacity $C_{ea}(\Phi,E)$~\cite{hw} (blue curve), and
entanglement-assisted quantum capacity $Q_{ea}(\Phi, E) = C_{ea}(\Phi,E)/2$ (green curve). Notice that for $k=1$ (noiseless case), $Q(\Phi,E)= C(\Phi,E)=Q_{ea}(\Phi,E)$. Also for $k^2\leq1/2$ the
channel becomes anti-degradable~\cite{cg,cgh} and the quantum capacity vanishes.
In the plot all the quantities have been rescaled by $g(E)$ (i.e. the optimal (entanglement-unassisted) transmission rate achievable for transmissivity $k=1$). }
\label{capacityplot}
\end{center}
\end{figure}

A plot of the above quantities is presented in Fig.~\ref{capacityplot} for the special case of the
attenuation channel~(\ref{param})  with $N_0=0$, for which the expressions simplify as follows
\begin{eqnarray}
C(\Phi ,E)&=&g(k^2 E) \;,  \nonumber \\
C_{ea}(\Phi ,E)&=&g(E) +g(k^2E) -g((1-k^2)E) = 2 Q_{ea}(\Phi,E)\;,\nonumber  \\
Q(\Phi,E)&=&\max \left\{0, g(k^2E) -g((1-k^2)E)  \right\}\;. \label{capplot}
\end{eqnarray}
Here $Q_{ea}(\Phi,E)$ is the {\it entanglement-assisted quantum capacity} of the channel $\Phi$ (i.e. the quantum capacity
which  is achievable when the sender and the receiver are provided with prior shared entanglement): from general results~\cite{bsst}
it is known to be always equal to half of the corresponding classical entanglement-assisted capacity. We also stress
 that for $N_0=0$ the reported value~(\ref{capplot})  for  $C(\Phi,E)$ is  the {\it exact} value of the classical capacity~\cite{gio}.

%%%%%%%%%%%%%%%%%%%%%%%%%%%%%%%%%%%%%%%%%%%%%%%%%%%%
%%%%%%%%%%%%%%%%%%%%%%%%%%%%%%%%%%%%%%%%%%%%%%%%%%

\subsection{The variety of quantum channel capacities}
\label{sectionVAR}

The three capacities defined in Eqs.~(\ref{coding}), (\ref{ceai}), (\ref{UBC})
 are related
as $Q(\Phi )\leq C(\Phi )\leq C_{ea}(\Phi )$ and
form a basis for defining
and investigating the diversity of various capacities of a quantum
communication channel, which arises by the application of additional
resources, such as reverse or direct communication, correlation, or
entanglement. In classical information theory it is well known that feedback does not
increase the Shannon capacity which is essentially the unique characteristic
of the classical channel. In the quantum case similar property is established~\cite{BOWENfeed} for the entanglement-assisted capacity $C_{ea}(\Phi)$. Regarding    the
quantum capacity $Q(\Phi)$, it is known that it can not be increased with an additional unlimited forward classical communication~~\cite{bsst,bns}.
However, $Q(\Phi)$ can be
increased if there is a possibility of transmitting the classical information
in the backward direction. Such a protocol  would allow one to create the
maximum entanglement between the input and output, which can be used for
quantum state teleportation. By this trick, even channels with zero quantum capacity
supplemented with a classical feedback can be used for the reliable transmission of
quantum information~\cite{nc,BOWENfeed} (a notable exception is PPT channels which have null  quantum capacity even in the presence of the feedback).
Furthermore Smith and Yard~\cite{SMITH} recently provided an explicit
example of an interesting phenomenon named~{\it superactivation}:
there exist cases in which, given two
quantum channels $\Phi_1$, $\Phi_2$ with zero quantum capacity
($Q(\Phi_1)=Q(\Phi_2)=0$), it is possible to have $Q(\Phi_1\otimes \Phi_2)>0$
(the latter being the capacity of the communication line $\Phi_1\otimes \Phi_2$ obtained by using
jointly $\Phi_1$ and $\Phi_2$). The example of~\cite{SMITH} was build by joining an anti-degradable channel $\Phi_1$ with
a PPT channel $\Phi_2$, and a more general construction was given in Ref.~\cite{BRANDAO}.

%%%%%%%%%%%%%%%%%%%%%%%%%%%%%%%%%%%%%%%%%%%%%%%%%%%%
\begin{figure}[t]
\begin{center}
\includegraphics[width=330pt]{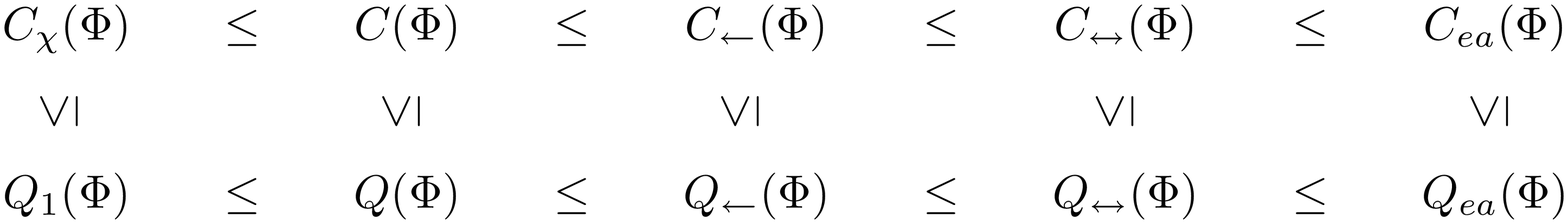}
\caption{Schematic comparison between the capacities of a memoryless quantum channel $\Phi$ (see text for the definitions of the various quantities).
The vertical ``inequalities" of the scheme follow from the fact  that each transferred  qubit of information can
carry a classical bit (e.g. see also Fig.~\ref{capacityplot}).}
\label{capacityscheme}
\end{center}
\end{figure}
%%%%%%%%%%%%%%%%%%%%%%%%%%%%%%%%%%%%%%%%%%%%%%%%%%

A comparison of  the various capacities which arise from employing additional
resources is presented in Fig.~\ref{capacityscheme},
where  the symbol ``$\leq$" should be understood as ``less than or equal to for all
channels and strictly less for some channels''~\cite{bdss}.
In such scheme $Q_{\leftarrow }$ and $C_{\leftarrow}$ denote respectively the quantum and classical capacities in the presence of a
feedback.  The symbol  $Q_{\leftrightarrow}$ represents the quantum capacity in the presence of two-side classical communication~\cite{eof}.
The corresponding classical capacity  $C_{\leftrightarrow}$ is computed under the limitation that the side communication is
independent of the message transmitted through the main channel~\cite{bdss}.
It is also known that
$C_{ea}=2Q_{ea}$~\cite{bsst} and that for some other pairs
both inequalities are possible. Further, one can construct so-called
``mother'' protocol which can implement all possible methods of
transmission, including those mentioned above when using various
additional resources (such such as feedback or entanglement)~\cite{adhv}.
We finally point out that of all the quantities
entering in scheme of Fig.~\ref{capacityscheme},  the function
$Q_1(\Phi)$ defined by the Eq.~(\ref{Q1}) is the only one which has no clear operational definition
in terms of information rate (apart from the cases in which it coincides with the quantum capacity -- e.g. for the degradable channels). Still we have inserted it in the list as it provides a lower bound for $Q(\Phi)$. Notice also that $C_\chi(\Phi)$ corresponds to the classical capacity of the channel $\Phi$
restricted to separable encodings.

In the classical information theory the role of the Shannon capacity is twofold: the converse statement of the coding theorem gives a fundamental upper bound on the channel performance which serves as an important benchmark for real information processing system. This is also the case for the present state of art with quantum channel capacities.
On the other hand, the direct statement of the coding theorem asserts that the bound is asymptotically achievable, although the proof
does not give a practical recipe. In fact for almost 50 years after the Shannon proof the real performance was well below the theoretical bound and only more recently efficient practical codes appeared with rates approaching the capacity. This is still to be done in the quantum case, and the existing quantum error-correcting codes are the first promising steps in that direction.  As an important example, in the papers \cite{GKP,HARRINGTON} symplectic codes were successfully applied to demonstrate constructively achievable rates close to the capacity for the additive classical noise channel (\ref{anc}).

\subsection{Relative entropy}
\label{relentropy}
Before concluding our  survey of quantum channel capacities  and their entropic expressions,
it is useful to consider
a fundamental quantity -- the quantum {\it relative entropy} -- which underlies many information
characteristics. Given two density operators $\rho,
\sigma$ in ${\cal H}$ ,  one defines the
relative entropy of $\rho$ with respect to $\sigma$ as
\begin{equation}\label{relent}
S(\rho \|\sigma ) \equiv  \tr \rho (\log \rho -\log \sigma )\;.
\end{equation}
This quantity provides an important (asymmetric) measure of
distinguishability of the  states $\rho$, $\sigma$; it is
nonnegative and is equal to zero
if and only if $\rho =\sigma$ ~\cite{wehrl,op}.

The most important  property of the relative entropy  which
underlies  several key facts in quantum information theory and nonequilibrium
statistical mechanics is {\it monotonicity} under the action of quantum
channels
\begin{eqnarray}\label{monot}
S(\rho \| \sigma )\geq S(\Phi [ \rho ]\|\Phi [ \sigma ])\;.
\end{eqnarray}
In words, this means that  states become less distinguishable after an irreversible
evolution $\Phi$. The first proof of the monotonicity property is due to
Lindblad \cite{lin75} who derived it from the strong
subadditivity of quantum entropy, Eq.~(\ref{ssa}).

To stress the relevance of the relative entropy in the context  of quantum information,
let us first notice that the quantum mutual information (\ref{qmu})
can be written as
\begin{equation}
I(\rho, \Phi)=S(\rho_{BR}\|\rho_{B}\otimes\rho_{R})\;,
\end{equation}
implying $I(\rho, \Phi)>0$ unless $\rho_{RB}=\rho_{R}\otimes\rho_{B}$ when $I(\rho, \Phi)=0$
(as already mentioned in Sec.~\ref{entexch}), and the data processing
inequality~(\ref{subqminfo1}) as a particular instance of Eq.~(\ref{monot}).

Next, the $\chi$-information~(\ref{chiq})
also admits a representation in terms of $S(\rho\|\sigma)$, namely~\cite{swoa}
\begin{equation} \label{chirel}
\chi \left( \{p _{x}\},\{\rho_{x}\}\right)=\sum_{x}p _{x}\; S\Big(\rho
_{x}\|\sum_{x'} p_{x'}\rho_{x'} \Big)\;,
\end{equation}
From  this expression
one can  invoke the monotonicity property~(\ref{monot}) to imply the following
data processing inequality
\begin{equation} \label{chirel1}
\chi \left( \{p _{x}\},\{\Phi(\rho_{x})\}\right) \leqslant
\chi \left( \{p _{x}\},\{\rho_{x}\}\right) \;,
\end{equation}
which in particular yields the bound (\ref{ebound}) when  $\Phi$ is taken  to be the q-c
channel transforming quantum
states into probability distributions
(in this case the $\chi$-function on the left hand side coincides
with the Shannon information of the measurement process associated with the q-c mapping).

The inequality~(\ref{chirel1}) has also profound implications for the coherent information.
Applying it to the identity (\ref{key2}) we obtain  that if the channel $\Phi $ is
anti-degradable, then for any state $\rho$
\begin{eqnarray}
I_{c}(\rho ,\Phi )\leq 0  \label{IC0}\;,
\end{eqnarray}
(correspondingly, for degradable
channel $I_{c}(\rho ,\Phi )\geq 0$). From the coding theorem (\ref{UBC}) it then follows that
{\it all anti-degradable channels have zero quantum capacity} as
anticipated at the end of Sec.~\ref{qcap}.

Another useful application of the monotonicity (\ref{monot})
concerns what may be called the {\it generalized H-theorem} which
states that a bistochastic (unital) evolution does not  decrease the entropy. In other words, the H-theorem implies that
given a unital channel $\Phi$  and arbitrary input state $\rho$, one has
\begin{eqnarray}
S(\Phi [ \rho ])\geq S(\rho )\;.\label{Hthm}
\end{eqnarray}
To see this in finite dimensional case ($\dim
\mathcal{H}=d<\infty$), use the identity
\begin{eqnarray}
S(\rho )=\log d-S(\rho \|I/d)\;,  \label{rel}
\end{eqnarray}
where $I/d$ is the chaotic state, and  notice that
\begin{eqnarray}
S(\Phi [ \rho ])&=&\log d-S(\Phi [ \rho ]\| I/d)=\log d-S(\Phi [ \rho ]\| \Phi[ I/d]) \nonumber \\
&\geqslant& \log d-S( \rho \|  I/d) = S(\rho)\;.
\end{eqnarray}
where  Eq.~(\ref{monot}) and the fact
that  $\Phi [ I/d] =I/d$ for bistochastic channels was used.

For a general channel $\Phi: \mathfrak{S} ({\cal H}) \rightarrow
\mathfrak{S} ({\cal H})$ an interesting characteristic is the
{\it minimal entropy gain} defined by the quantity
\begin{eqnarray}\label{gainfin}
G(\Phi )=\inf_{\rho }   \; \Big[ S(\Phi [ \rho ])-S(\rho )\Big] \;,
\end{eqnarray}
In contrasts to nonadditivity of the other similar quantity --
the minimal output entropy $\inf_\rho S(\Phi[\rho])$ -- which we introduced at the end of
Sec.~\ref{partII}
the quantity $G(\Phi )$
is additive with respect to tensor
product of channels (i.e.
$G(\Phi_1\otimes\Phi_2)=G(\Phi_1)+G(\Phi_2)$)
as a simple consequence of  the strong subadditivity (\ref{ssa}), see~\cite{al}.
For finite dimensional system  it is easy to see that
\begin{eqnarray}
-\log_2 d\leq G(\Phi )\leq 0 \;, \label{2}
\end{eqnarray}
(this  follows directly from the fact that von Neumann entropies
are upperbounded by $\log_2 d$).
However, as shown in~\cite{heg}, there is much better lower estimate
\begin{eqnarray}
-\log_2 \Vert \Phi [ I]\Vert \leq G(\Phi )\;.  \label{14}
\end{eqnarray}
which holds also for infinite dimensional systems assuming that the channel $\Phi$ is such that $\Phi [
I]\equiv\sum_{j=1}^{\infty }V_{j}V_{j}^{\dag}$ is a bounded operator. This
implies that the generalized H-theorem is valid also for infinite dimensional
unital evolutions $\Phi$ if one restricts to the input states $\rho$ with finite entropy.
The inequality $G(\Phi )\leq 0$ no longer holds; for example, if $\Phi$
is Bosonic Gaussian channel with parameters $(K, l, \beta), \det K\neq 0$ (see Sec. \ref{gpgc}) then the minimal entropy gain is equal to
\begin{eqnarray}
G(\Phi )=\log_2 |\det K|  \label{INF1}\;,
\end{eqnarray}
and is attained on Gaussian states \cite{heg}. The quantity $|\det
K|$ is the coefficient of the change of the classical phase space
volume under the linear transformation $K$ (which can take arbitrary positive values).
These results also suggest that for a
general irreversible quantum evolution the role of this coefficient is
played by the quantity $\Vert \Phi [ I]\Vert ^{-1}$.

\section{Summary and outlook}\label{sec:conclusion}

In this review we considered evolutions of open quantum systems from a quantum information viewpoint.
More specifically, we discussed several scenarios where a sender and a receiver establish
communication line by using some physical degrees of freedom (the information carrier) subject to quantum noise from the environment.
The resulting transformation of sender's input states into receiver's output states -- the quantum channel in the Schr\"{o}dinger picture-- is described
by a completely positive trace preserving (CPTP) map.
Several alternative representations of such maps were introduced and their main features were
analyzed.
In particular we have reviewed the operator-sum (Kraus) and unitary/isometric   representations of a quantum channel,
as well as its characterization in terms of
the corresponding Choi-Jamiolkowski state.  Composition rules along with the notions of dual (Heisenberg picture) and complementary channels have been presented.
The general treatment was supplied with discussion of important particular cases of qubit, depolarizing, erasure channels as well as entanglement-breaking and dephasing channels.
A whole section was dedicated to Bosonic Gaussian channels which  constitute a basic class of information processors for continuous variable systems and
provide a representation for some of the most usable quantum communication protocols.

In the second part of the review we surveyed approaches to evaluation of the quality of a given quantum channel,
basing on the notions of channel capacities and their entropic expressions. For the sake of clarity, we restricted the analysis to the basic case of memoryless communication models, where the noise  operates identically and independently on each of the sender's inputs.
We started with recollections of the Shannon entropy of a classical  random source and  the Shannon capacity of a classical channel (relevant for a communication line where all the stages of the information transferring --  coding, transmission, decoding -- are treated in terms of classical random processes), explaining how these quantities
acquire operational meaning in the asymptotic of very long messages.
Moving to the quantum domain, we demonstrated  how these quantities
admit multiple generalizations. Several different notions of channel capacities have to be introduced
in view of the fact that  a quantum communication line can be used to
transfer either classical or quantum messages, and assisted by various additional resources such as
entanglement shared between the communicating parties or the classical feedback, which in the Shannon theory either do not exist (entanglement) or do not increase the capacity (feedback). These quantities have operational definitions generalizing that of the Shannon capacity and admit closed expressions in entropic terms given by fundamental quantum coding theorems. The quantum correlations (entanglement) display themselves in the increase of information transmission rates as compared to protocols without entanglement.
The notable cases of this phenomenon discussed in our paper include the superadditivity of the classical information transmission rates with respect to entangled decodings and encodings, the gain of the input-output entanglement assistance and superactivation of zero quantum channel capacity.

Information-theoretic view opens thus a completely new perspective of quantum irreversible evolutions -- noisy communication channels -- and enlightens fascinating landscape of the channel entropic characteristics, in which entanglement plays a crucial role.

\ack VG acknowledges support from  MIUR through FIRB-IDEAS Project
No. RBID08B3FM. AH was partially supported by RFBR grant and the RAS
program ``Dynamical systems and control theory''.

 \end{document}